\documentclass[12pt]{article}
\pdfoutput=1
\usepackage[latin1]{inputenc}

\usepackage{amsmath}
\usepackage{amsfonts}
\usepackage{amssymb}
\usepackage{graphicx}
\usepackage{geometry}
\usepackage{amssymb,epsfig}
\usepackage{hyperref}
\usepackage{subfig}
\usepackage{slashed}
\usepackage{comment}

\makeatletter
\renewcommand\section{\@startsection {section}{1}{\z@}%
                                 {-3.5ex \@plus -1ex \@minus -.2ex}
                                   {2.3ex \@plus.2ex}%
                                   {\normalfont\large\bfseries}}
\renewcommand\subsection{\@startsection{subsection}{2}{\z@}%
                                   {-3.25ex\@plus -1ex \@minus -.2ex}%
                                     {1.5ex \@plus .2ex}%
                                     {\normalfont\bfseries}}
\renewcommand\subsubsection{\@startsection{subsubsection}{3}{\z@}%
                                   {-3.25ex\@plus -1ex \@minus -.2ex}%
                                     {1.5ex \@plus .2ex}%
                                     {\normalfont\itshape}}
\makeatother

\def\pplogo{\vbox{\kern-\headheight\kern -29pt
\halign{##&##\hfil\cr&{\ppnumber}\cr\rule{0pt}{2.5ex}&\ppdate\cr}}}
\makeatletter
\def\ps@firstpage{\ps@empty \def\@oddhead{\hss\pplogo}%
  \let\@evenhead\@oddhead 
}
\def\maketitle{\par
 \begingroup
 \def\thefootnote{\fnsymbol{footnote}}
 \def\@makefnmark{\hbox{$^{\@thefnmark}$\hss}}
 \if@twocolumn
 \twocolumn[\@maketitle]
 \else \newpage
 \global\@topnum\z@ \@maketitle \fi\thispagestyle{firstpage}\@thanks
 \endgroup
 \setcounter{footnote}{0}
 \let\maketitle\relax
 \let\@maketitle\relax
 \gdef\@thanks{}\gdef\@author{}\gdef\@title{}\let\thanks\relax}
\makeatother

\numberwithin{equation}{section}

\renewcommand{\dag}{\dagger}

\newcommand{\be}{\begin{equation}}
\newcommand{\bea}{\begin{eqnarray}}
\newcommand{\ee}{\end{equation}}
\newcommand{\eea}{\end{eqnarray}}
\newcommand\beq{\begin{equation}}
\newcommand\eeq{\end{equation}}

\newcommand{\mc}{\mathcal}

\renewcommand{\t}{\tilde}

\newcommand{\ba}{\begin{align}}
\newcommand{\ea}{\end{align}}
\newcommand{\bg}{\begin{gather}}
\newcommand{\eg}{\end{gather}}
\newcommand{\bseq}{\begin{subequations}}
\newcommand{\eseq}{\end{subequations}}

\renewcommand{\tanh}{\mathop{\rm th}\nolimits}

\begin{document}

\setcounter{page}0
\def\ppnumber{\vbox{\baselineskip14pt
}}
\def\ppdate{\footnotesize{SU-ITP-14/01}} \date{}

\author{Anson Hook$^1$, Shamit Kachru$^{2,3}$, Gonzalo Torroba$^2$ and Huajia Wang$^2$\\
[7mm]
{\normalsize \it $^1$ School of Natural Sciences, Institute for Advanced Study }\\
{\normalsize \it Princeton, NJ 08540, USA}\\
[7mm]
{\normalsize \it $^2$ Stanford Institute for Theoretical Physics, Stanford University }\\
{\normalsize \it Stanford, CA 94305, USA}\\
[7mm]
{\normalsize \it $^3$ Theory Group, SLAC National Accelerator Laboratory}\\
{\normalsize \it Menlo Park, CA 94309, USA}\\
}

\bigskip
\title{\bf  Emergent Fermi surfaces, fractionalization  \\ and duality  in supersymmetric QED
\vskip 0.5cm}
\maketitle

\begin{abstract}
We study the physics of 3d supersymmetric abelian gauge theories (with small supersymmetry breaking
perturbations) at finite density.  Using mirror symmetry,
which provides a natural generalization of the duality between the XY model and the abelian Higgs model
but now including fermionic fields,  we see many dynamical phenomena conjectured to be of
relevance in condensed matter systems.  In particular, we find examples of the emergence of a Fermi surface at low
energies from hybridization of fermions localized at magnetic defects at high energies, as well as
fractionalization of charged fermions into spinon-holon pairs with the concomitant appearance of emergent gauge
fields.  We also find dual descriptions for Fermi surfaces coupled to critical bosons, which give
rise to non-Fermi liquids.

\end{abstract}
\bigskip
\newpage

\addtocontents{toc}{\protect\setcounter{tocdepth}{2}}

\tableofcontents

\vskip 1cm

\noindent

\vspace{0.5cm}  \hrule
\def\thefootnote{\arabic{footnote}}
\setcounter{footnote}{0}

\section{Introduction}\label{sec:intro}

Strongly coupled systems at finite density display fascinating phenomena which are not seen in their zero density cousins. Examples abound across different areas, including strange metals and high temperature superconductors, the QCD quark-gluon plasma, fractional quantum Hall states, and
many others. Understanding the physics responsible for these phases of matter is, however, extremely challenging, and it is crucial to develop new analytic tools for this purpose. 

Supersymmetry provides a tool that is well known to provide 
theoretical control over highly non-trivial dynamics.
In this work, and in~\cite{Hook:2013yda}, we propose to study finite density systems using quantum field theory (QFT) dualities of supersymmetric gauge theories.
These dualities can provide analytic control even over situations with strong coupling, where
non-perturbative physics plays an important role.
The lessons learned from such supersymmetric theories and their dualities may be useful for understanding more general QFTs.
Indeed, the supersymmetric models under study can include fairly mild generalizations of the conventional
field theories of interest in condensed matter physics, such as supersymmetric abelian gauge theories
with charged ``electron" flavors (as well as their scalar superpartners) -- this point has been
discussed also in \cite{Sachdev:2008wv}.
However, the dualities of supersymmetric gauge theories and other SUSY results have not yet been systematically studied at finite density. There may be different reasons for this.  One reason is that supersymmetry does not seem to be 
directly realized in any condensed matter systems of interest. This situation may now be changing -- see ~\cite{Lee:2006if,Roy:2012wz,Grover:2013rc} for proposals which could realize supersymmetric field theories in condensed matter laboratories.  Another reason, from a more theoretical perspective, is that finite density generically breaks SUSY explicitly, and so a priori it is not clear whether
the techniques of supersymmetry can help one understand phenomena of interest in field theory
at finite density.

Our goal in this paper is to show that SUSY dualities constitute a powerful tool for analyzing strong dynamics at finite density. In particular, we will see that they can controllably realize many phenomena which are believed to occur in strongly correlated systems, such as fractionalization, emergent gauge fields and
novel states of compressible quantum matter.
As in the relativistic setup, we hope that these results will be relevant for understanding useful
toy models of condensed matter systems.

In~\cite{Hook:2013yda} it was shown that some fraction of SUSY can be preserved even at finite density. This allows one to apply exact SUSY results to finite densities of BPS defects. Here we will explore another avenue. We will add a finite density of nonsupersymmetric impurities to a SUSY theory that admits a dual description, and will work in a regime where these sources are sufficiently small (compared to some physical scale that characterizes the SUSY duality, as we explain in more detail below). The duality deduced from the supersymmetric limit then holds to a good approximation and, as we shall see, can lead to striking predictions for the low energy dynamics.

In this work we study three-dimensional SUSY QED (SQED) in the presence of magnetic impurities. This theory exhibits a type of particle-vortex duality called ``mirror symmetry'' (reviewed in \S \ref{sec:N4}) that turns out to be extremely useful for analyzing the IR physics at finite density. This is a SUSY cousin of the abelian Higgs model/XY model duality~\cite{Peskin,Dasgupta}. The addition of supersymmetry leads to various novelties and allows for a very precise mapping between the dual descriptions. Our main results will be explicit realizations within SQED of phenomena that are conjectured in strongly interacting quantum matter: fractionalization of electrons, emergent gauge fields in 2+1 dimensions, and compressible phases of bosons (``Bose metals'' \cite{bose}). Mirror symmetry provides an analytic handle on these effects in the limit of strong coupling.

We begin in \S \ref{sec:N4} with the simplest case of $U(1)$ SQED with $\mc N=4$ SUSY and $N_f=1$ electron hypermultiplet. Mirror symmetry relates the IR dynamics to the theory of a single free vortex hypermultiplet. We add a density of external magnetic impurities to the UV theory and find that this maps to a chemical potential for vortices. Mirror symmetry reveals an emergent Fermi surface in the IR: the low energy phase is found to be a Fermi liquid of vortices. 
This can be viewed as a simple and controlled example of hybridization -- fermion zero modes
localized at the magnetic impurities in the electric theory, have been liberated and form a Fermi liquid in the magnetic dual.
We also discuss the mapping of nonsupersymmetric deformations in the electric and magnetic theory, and use them to stabilize a superfluid instability of the theory.

Next, in \S \ref{sec:N4gen} we analyze SQED with $N_f>1$ flavors. The mirror dual now has a `magnetic' gauge group $U(1)^{N_f-1}$ with $N_f$ charged vortices. This provides an analytically controlled realization of emergent gauge fields and fractionalization of electrons. Unlike the $N_f=1$ case, now both the electric and magnetic descriptions are strongly interacting in the IR. Nevertheless, the nontrivial mapping of global symmetries still allows us to derive useful results. In particular, using mirror symmetry for the self-dual $N_f=2$ theory, we argue that the theory of a gauge field interacting with a Fermi surface is dual to an abelian gauge theory with a lattice of external magnetic impurities. 
This provides a new window on theories similar to those which occur in discussions of non-Fermi liquid phases and strange metals in heavy fermion systems and exotic superconductors.

Finally, in \S \ref{sec:N2} we discuss duality for $\mc N=2$ SQED doped with magnetic impurities. The $N_f=1$ model has richer dynamics than its $\mc N=4$ counterpart, with a mirror dual that is a supersymmetric generalization of the strongly interacting Wilson-Fisher fixed point. At tree level we  find again a vortex Fermi liquid, but interactions with a gapless boson lead to non-Fermi liquid behavior. We will discuss its one loop properties, as well as deformations that can lead to a controlled expansion at low energies.  \S \ref{sec:concl} presents our conclusions and future directions.

\section{Emergent Fermi liquid in $\mc N=4$ SQED with one flavor}\label{sec:N4}

In this section we will show that the IR dynamics of 3d $\mc N=4$ SQED doped with magnetic impurities gives rise to an emergent Fermi surface corresponding to a Landau-Fermi liquid of fermionic vortices. The dynamics in the presence of sources that preserve 1/2 of the supercharges was studied in~\cite{Hook:2013yda}; here we will instead focus on nonsupersymmetric defects.
First we review the matter content, dynamics and duality of the theory. Next we turn on a chemical potential for the topological $U(1)_J$ symmetry, which arises from external vortices. We discuss the consequences of this in the UV (electric) theory and then use mirror symmetry to analyze the low energy dynamics, where we will find an emergent Fermi surface for the composite fermions.

\subsection{Super QED and mirror symmetry}

We begin by reviewing the dynamics of 3d SQED with $U(1)$ gauge group, one flavor and $\mc N=4$ supersymmetry. At strong coupling, the theory flows to a nontrivial conformal field theory that admits a mirror dual description in terms of free vortices with the same amount of supersymmetry~\cite{Mirror,deBoer:1996mp,Kapustin:1999ha}. We will explain in detail the mapping of external sources and global symmetries, which will be important for adding electric and magnetic impurities to the theory.

Let us recall first the fact that in $2+1$ dimensions a gauge field $A_\mu$ is dual to a scalar $\gamma$ --the ``dual photon''. The duality transformation is
\be\label{eq:Fdual}
F_{\mu\nu}= \epsilon_{\mu\nu\rho} \partial^\rho \gamma\,.
\ee
The abelian gauge field gives rise to a global $U(1)_J$ current that shifts the dual photon,
\be\label{eq:Jtop}
J_\mu= \frac{1}{2} \epsilon_{\mu\nu\rho} F^{\nu\rho}= \partial_\mu \gamma\,.
\ee
Note that current conservation follows from the Bianchi identity for the gauge field.
The $U(1)_J$ current plays an important role in particle/vortex dualities, because its charge is carried by sources of magnetic flux, i.e. vortices or monopoles.

\subsubsection{The electric description}

In the UV, the theory is weakly coupled and described in terms of the electric variables.
It contains: {\textit i)} an $\mc N=4$ vector multiplet $\mc V$, which consists of an $\mc N=2$ vector multiplet $V=(A_\mu,\,\sigma,\, \lambda, D)$ and a chiral multiplet $\Phi = (\phi, \psi_\phi, F_\phi)$; and \textit{ ii)} an $\mc N=4$ hypermultiplet $\mc Q$, consisting of two $\mc N=2$ chiral multiplets $Q= (q, \psi_q, F_q)$ and $\t Q= (\t q, \psi_{\t q}, F_{\t q})$ of opposite charge under the $U(1)$. The global symmetries are $SU(2)_L \times SU(2)_R \times U(1)_J$, where the nonabelian subgroups are R-symmetries, and $U(1)_J$ is the topological symmetry introduced in (\ref{eq:Jtop}). Since the global symmetries are not manifest when the theory is written in terms of $\mc N=2$ superfields, it is instead more convenient to group the fields as follows:
\begin{center}
\be
\begin{tabular}{c|ccc}
&$SU(2)_L$&$SU(2)_R$&$U(1)_J$\\
\hline
&&&\\[-12pt]
$e^{2\pi i \gamma/g^2}$  & 1& 1 & 1  \\
&&&\\[-12pt]
$\phi_{(ij)}$  & 3& 1 & 0  \\
&&&\\[-12pt]
$\lambda_{ia}$  & 2& 2 &  0 \\
&&&\\[-12pt]
$q_a$  & 1& 2 & 0  \\
&&&\\[-12pt]
$\psi_i$  & 2& 1 & 0 
\end{tabular}
\ee
\end{center}
Here $\gamma$ is the dual photon, which is the only field charged under the topological $U(1)_J$. The triplet $\phi_{(ij)}= (\sigma, {\rm Re}\,\phi,\,{\rm Im}\,\phi)$. All the fermions are two-component (i.e. 3d Dirac fermions).
The gauginos $\lambda_{ia}$ group the partner $\lambda$ of $A_\mu$ and $\psi_\phi$; they satisfy a reality condition $\lambda^\dag_{ia}= \epsilon_{ij} \epsilon_{ab} \,\lambda_{jb}$. The doublet of complex scalars is defined as $q_a = (q , \t q^*)$, and for the fermions, $\psi_i = (\psi_q,\,\psi_{\t q}^*)$. The notation for all the fields here is the same as in~\cite{Hook:2013yda}.

The Lagrangian of the theory is
\be
L_\text{el}= L_V(\mathcal V) + L_H(\mathcal Q, \mathcal V)
\ee
where the kinetic terms for the vector superfield are
\be
L_V(\mathcal V) =\frac{1}{g^2} \left(- \frac{1}{4} F_{\mu\nu}^2 + \frac{1}{2} (\partial_\mu \phi_{(ij)})^2+ i \bar \lambda_{ia} \not \! \partial \lambda_{ia}+ \frac{1}{2} D_{(ab)}^2 \right)
\ee
and the terms for the hypermultiplet are given by
\be
L_H(\mathcal Q, \mathcal V)= |D_\mu q_a|^2 + i \bar \psi_i \not \! \!D \psi_i- \phi_{(ij)}^2 |q_a|^2- \phi_{(ij)} \bar \psi_i \psi_j + \sqrt{2}(i \bar \lambda_{ia} q^\dag_a \psi_i + c.c.)+ D_{(ab)} q_a^\dag q_b
\ee
Here $D_{(ab)}$ is the triplet $(D, {\rm Re}\,F,{\rm Im} \,F)$ of auxiliary components of the $\mc N=4$ vector superfield. 

The theory has a Coulomb branch parametrized by $(\phi_{(ij)}, \,\gamma)$, and for $N_f=1$ there is no Higgs branch (the only solution to the F and D-term conditions is $q=\t q=0$).
There is only one coupling $g^2$ with dimensions of mass.  In the IR, the theory with $N_f=1$
flows to a fixed point that admits a weakly coupled dual, while the $N_f > 1$ theories (studied in \S \ref{sec:N4gen}) have a non-trivial superconformal field theory living at the origin of the moduli space in both descriptions.

\subsubsection{Mirror symmetry and magnetic description}

The IR fixed point admits a description in terms of weakly coupled magnetic variables, which are the vortices of the original theory. The dual arises as the sigma model along the Coulomb branch. Integrating out the massive hypermultiplet at an arbitrary Coulomb branch point $\vec \phi$ corrects the gauge coupling function
\be\label{eq:grunning}
\frac{1}{g_L^2}= \frac{1}{g^2}+ \frac{1}{4\pi |\vec \phi|}\,.
\ee
This expression is exact to all orders in perturbation theory and also nonperturbatively. The effective theory is then a sigma model with Taub-NUT metric~\cite{Mirror,deBoer:1996mp}, 
\be\label{eq:Taub-NUT1}
L_\text{eff}= \frac{1}{2g^2} \left(H(\phi)(\partial_\mu \vec \phi)^2 + H^{-1}(\phi) (\partial_\mu \gamma + \frac{1}{2\pi}\vec \omega \cdot \partial_\mu \vec \phi)^2 \right)
\ee
with
\be\label{eq:Taub-NUT2}
H(\phi) = 1 + \frac{g^2}{4\pi | \vec \phi|}\;\;,\;\;\vec \nabla \times \omega = \vec \nabla H\,.
\ee

In the IR limit $g^2/|\phi| \to \infty$ we obtain the mirror dual, which consists of a free $\mc N=4$ hypermultiplet $(V_+, V_-)$. The bosons are obtained from the map
\be\label{eq:quantum-mapb}
v_i \equiv \left(
\begin{matrix}
v_+ \\ v_-^*
\end{matrix}\right)= \sqrt{\frac{|\vec \phi|}{2\pi}} e^{2\pi i \gamma/g^2}
\left(
\begin{matrix}
\cos \frac{\theta}{2} \\
e^{i \varphi} \sin \frac{\theta}{2}
\end{matrix}
\right)\;\;,\;\;\vec \phi = |\vec \phi| (\cos \theta, \sin \theta \cos \varphi, \sin \theta \sin \varphi)
\ee
and the fermions
\be\label{eq:quantum-mapf}
\psi_a \equiv \left(
\begin{matrix}
\psi_+ \\ \psi_-^*
\end{matrix}\right)= \frac{1}{\sqrt{2}}\,\frac{v_a \lambda}{2\pi\,\sum_i|v_i|^2}\,.
\ee
Here $v_i=(v_+,\,v_-^*)$ is a doublet of complex scalars, and $\psi_a=(\psi_+,\,\psi_-^*)$ is a doublet of 3d Dirac fermions. (We don't distinguish between the $2$ and $\bar 2$ representations). 
See also the recent discussion in~\cite{Sachdev:2008wv}, where the fermionic terms are obtained using supersymmetry transformations.

The global symmetries of the theory are
\begin{center}
\be\label{table:magnetic}
\begin{tabular}{c|ccc}
&$SU(2)_L$&$SU(2)_R$&$U(1)_J$\\
\hline
&&&\\[-12pt]
$v_i$  & 2& 1 & 1  \\
&&&\\[-12pt]
$\psi_a$  & 1& 2 & 1
\end{tabular}
\ee
\end{center}
The Lagrangian is simply
\be
L_\text{mag}= |\partial_\mu v_i|^2 + i \bar \psi_a  \not \! \partial \psi_a\,.
\ee 
Note that at low energies the scalars and fermions are approximately decoupled. There are interactions suppressed by powers of $|\vec \phi|/g^2$, which can be obtained by using the exact one loop action (\ref{eq:Taub-NUT1}).

We see that the gauginos of the $U(1)$ electric theory (which are gauge singlets) bind with the dual photon $e^{2\pi i \gamma/g^2}$ (and with a combination of Coulomb branch scalars determined by the Taub-NUT geometry) to yield the elementary charged fermions $\psi_a$ of (\ref{eq:quantum-mapf}). We will shortly show how to obtain an emergent Fermi surface for these composite fermions. We also note that, consistently with the $\mc N=4$ supersymmetry, we also obtain elementary scalars $v_i$ charged under the $U(1)_J$. The Coulomb branch of the electric theory maps to the Higgs branch of the magnetic description via (\ref{eq:quantum-mapb}).
Below we will use soft supersymmetry breaking deformations to lift these directions.

Mirror symmetry is a supersymmetric version of the XY/abelian Higgs duality of~\cite{Peskin,Dasgupta}, but there are also differences that will be crucial for us. The high degree of supersymmetry allows to determine explicitly the map between the electric and magnetic variables, and provides a fully calculable IR description. More importantly for our purpose, there are composite fermions in the IR, which are absent from the nonsupersymmetric version of the duality. It is in terms of these fermions that we will obtain the emergent Fermi surface.

\subsubsection{External sources}

In this work we will deform $\mc N=4$ SQED and its mirror dual by adding external sources, so let us now discuss how they map across the duality.

In the electric theory, we turn on a background vector multiplet $\hat{\mc V}$ that couples to the $U(1)_J$ supercurrent. In terms of the dynamical vector multiplet ${\cal V}$, these sources give BF interactions
\be\label{eq:LBF}
L_{BF}= \frac{1}{2\pi}\left(- \frac{1}{2} \epsilon^{\mu\nu\rho} A_\mu \hat F_{\nu\rho}+ \phi_{(ij)} \hat D_{(ij)}+ \hat \phi_{(ab)} D_{(ab)} + i \lambda_{ia} \hat \lambda_{ia}\right)\,.
\ee
Here $D_{(ab)}$ is the triplet $(D, {\rm Re}\,F,{\rm Im} \,F)$ of auxiliary components of the $\mc N=4$ vector superfield.  The supersymmetric extension of this includes the triplet $\hat D_{(ij)}$, which gives real and complex masses to the hypermultiplet, and $\hat \phi_{(ab)}$ includes the usual (real) FI term plus a complex linear source for the F-term.

As is clear from $L_{BF}$, the sources transform according to
\begin{center}
\be
\begin{tabular}{c|ccc}
&$SU(2)_L$&$SU(2)_R$&$U(1)_J$\\
\hline
&&&\\[-12pt]
$\hat A_\mu$  & 1& 1 & 0  \\
&&&\\[-12pt]
$\hat D_{(ij)}$  & 3& 1 & 0\\
&&&\\[-12pt]
$\hat \phi_{(ab)}$  & 1& 3 & 0\\
&&&\\[-12pt]
$\hat\lambda_{ia}$  & 2& 2 & 0
\end{tabular}
\ee
\end{center}
Note that integrating out the auxiliary $D_{(ab)}$ sets
\be
D_{(ab)}= - g^2 \left(q_a^\dag q_b+ \frac{1}{2\pi}\hat \phi_{(ab)}\right)
\ee
and
gives a quartic interaction
\be
V_D= \frac{1}{2} g^2  \left(q_a^\dag q_b+ \frac{1}{2\pi}\hat \phi_{(ab)}\right)^2\,.
\ee

The action for the magnetic theory in the presence of sources is given by 
\be
L_H(\hat{\mc Q}, \hat {\mc V})=|\hat D_\mu v_i|^2 + i \bar \psi_a \not \! \!\hat D \psi_a- \hat \phi_{(ab)}^2 |v_i|^2- \hat \phi_{(ab)} \bar \psi_a \psi_b + \sqrt{2}(i \bar {\hat \lambda}_{ia} v^\dag_i \psi_a + c.c.)+ \hat D_{(ij)} v_i^\dag v_j
\ee
where $\hat A_\mu$ is the $U(1)_J$ connection
\be
\hat D_\mu = \partial_\mu - i  \hat A_\mu\,.
\ee
$\hat D_{(ij)}$ is a background D-term for $v_i^\dag v_j$, and $\hat \phi_{(ab)}$ gives real and complex masses.

\subsection{Adding a chemical potential for the $U(1)_J$ symmetry}\label{subsec:N4el1}

Now we come to the main part of our analysis, where we add a constant chemical potential $\mu$ for the $U(1)_J$ symmetry. This is an expectation value for the background gauge field $\hat A_0 = \mu$. In order to be able to use the mirror duality, the source is required to be small in units of the gauge coupling,
\be\label{eq:smallA0}
\hat A_0 \ll g^2\,.
\ee
In the UV only the dual photon is charged under this symmetry; in particular, there are no charged fermionic excitations. We will show that an emergent Fermi surface of the fermions $\psi_a$ forms at low energies.

Let us begin by analyzing the weakly coupled limit in terms of the electric variables. Physically, the chemical potential for $U(1)_J$ comes from inserting a lattice of nondynamical vortices that source a magnetic field for the gauge $U(1)$. We will work at distances much larger than the lattice spacing, so that the magnetic flux from the vortices can be averaged and the $U(1)_J$ chemical potential is approximately constant.\footnote{We note that the structure of the lattice could be important for transport properties, something that would be interesting to study in more detail.}
For a constant chemical potential, neglecting the interactions with the hypermultiplet (which is justified e.g. far along the Coulomb branch due to the large hypermultiplet mass), one obtains
\be\label{eq:dynamicB}
F_{ij} = \epsilon_{ij} \frac{g^2}{2\pi} \hat A_0\,.
\ee
The limit where the chemical potential is exactly constant is a bit ambiguous because any constant $F_{ij}$ is a solution. Here, we determine (\ref{eq:dynamicB}) by allowing for a small position dependence and requiring that $F_{ij} \to 0$ in the absence of sources. At this order, we then have a background magnetic field for the gauge $U(1)$. The total charge (\ref{eq:Jtop}) carried by this configuration is
\be
Q = \int d^2 x J_0 = \frac{g^2}{2\pi} \hat A_0\,V_2
\ee
where $V_2$ is the spatial volume. We can interpret this physically as having inserted a lattice of nondynamical vortices with density $ \frac{g^2}{2\pi} \hat A_0$, each carrying a unit of magnetic flux.

Classically, the nonzero magnetic field $B= \frac{1}{2} \epsilon_{ij} F_{ij}$ gives Landau levels for the hypermultiplet bosons and fermions, with a nonsupersymmetric spectrum
\be\label{eq:LL}
E_\text{fermion}^{(n)}= \pm \sqrt{2n |B|}\;,\;E_\text{boson}^{(n)}= \pm \sqrt{(2n+1) |B|}\,,
\ee
with $n=0,\,1,\,\ldots$ the Landau level index. Each level has degeneracy
\be
\nu = \frac{|B|}{2\pi}\,V_2\,.
\ee
The fermion has a lowest Landau level with $E=0$, but the bosonic spectrum is gapped. Also, the $n \ge 1$ levels have both particle and antiparticle solutions, while the $n=0$ level is special in that it only describes particles. When the representations are combined into Dirac fermions, the degeneracy of the $n \ge 1$ levels is then twice that of the lowest Landau level, something that can lead to interesting experimental consequences in systems such as graphene.
At tree level, the Coulomb branch of the electric theory is not lifted.

The splitting between bosons and fermions leads to quantum-mechanical corrections to this picture. The electric description is useful far along the Coulomb branch, $|\vec \phi | \gg g^2$, where the theory is weakly coupled; so let us see what happens in this limit. We need to calculate the Coleman-Weinberg potential along the Coulomb branch, from integrating out the massive hypermultiplet with spectrum (\ref{eq:LL}). The result can be deduced from the following argument. The Taub-NUT sigma model (\ref{eq:Taub-NUT1}) is obtained by integrating out the hypermultiplets at an arbitrary Coulomb branch coordinate $\vec \phi$, and keeping terms to second order in derivatives. Now, recalling (\ref{eq:smallA0}), far along the Coulomb branch we have $|B| \ll |\vec \phi|^2$, so we only need the leading dependence on the magnetic field. This dependence is quadratic and is captured by the $F_{\mu\nu}^2$ term of the Taub-NUT one loop action,
\be
L = -\frac{1}{4g^2} \frac{1}{1+ \frac{g^2}{4\pi | \vec \phi|}} F_{\mu\nu}^2 + \ldots
\ee
This implies that the Coleman-Weinberg potential in a nonzero magnetic field (\ref{eq:dynamicB}), far along the Coulomb branch, is given by
\be\label{eq:CWapproxUV}
V_\text{CW} \approx \frac{B^2}{16 \pi | \vec \phi|}\,.
\ee

\begin{figure}[h!]
\begin{center}
\includegraphics[width=0.6\textwidth]{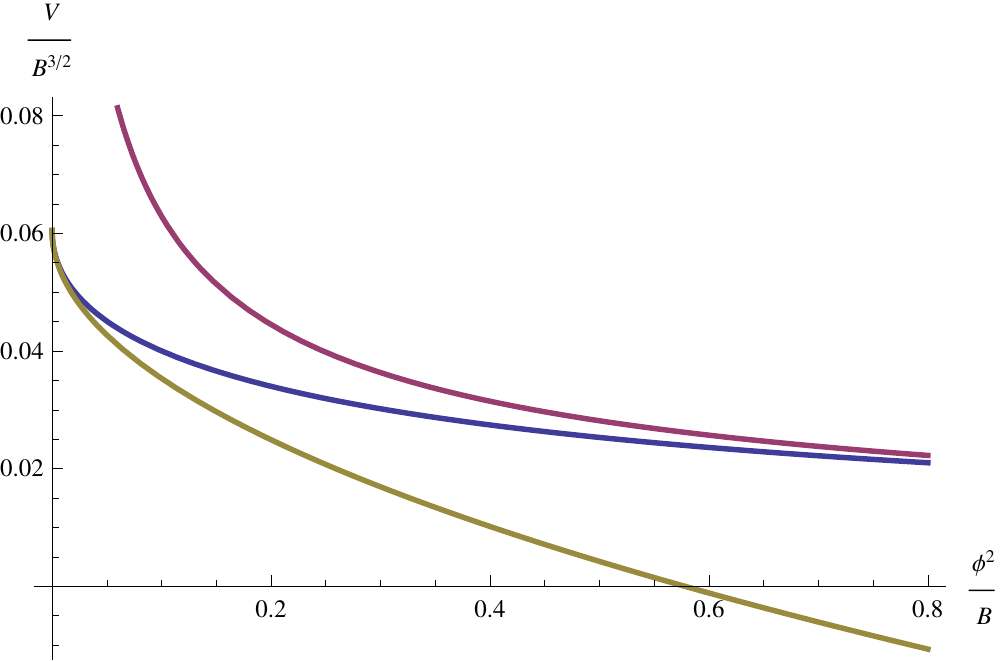}
\end{center}
\caption{\small{Plot of the exact potential (\ref{eq:CWexact}) [in blue] and the approximations at large values (\ref{eq:CWapproxUV}) [in violet] and small values  (\ref{eq:CWapproxIR}) [in brown], that produce the instability along the Coulomb branch.}}\label{fig:CWplot}
\end{figure}

This agrees with the limit of large $| \vec \phi|$ of~\cite{Cangemi:1994by}, who also calculated the exact one-loop answer,
\be\label{eq:CWexact}
V_\text{CW}= \left(\frac{|B|}{4\pi} \right)^{3/2}\,\int_0^\infty ds \,\frac{e^{- |\vec \phi|^2 s/|B|}}{s^{3/2}}\,\tanh \frac{s}{2}\,.
\ee
This integral may be expressed in terms of the generalized Riemann-zeta function $\xi(z, a)$, in the following form:
\be\label{eq:CWfinal}
V_\text{CW}=\frac{|B|^{3/2}}{\sqrt{2}\pi}\left\{\xi\left(-\frac{1}{2}, \frac{|\vec\phi |^2}{2|B|}+\frac{1}{2}\right)-\frac{1}{2}\,\xi\left(-\frac{1}{2}, \frac{|\vec\phi |^2}{2|B|}\right)-\frac{1}{2}\,\xi\left(-\frac{1}{2}, \frac{|\vec\phi |^2}{2|B|}+1\right)\right\}\,.
\ee
Using this result we can derive an analytic expression valid near the origin:
\be\label{eq:CWapproxIR}
V_\text{CW} \approx\frac{\left(2 \sqrt{2}-1\right)\zeta(3/2)}{8 \pi ^2} |B|^{3/2}-\frac{|B|}{4 \pi }\,|\vec \phi|\,.
\ee
The exact expression  (\ref{eq:CWexact}) and the series expansions  (\ref{eq:CWapproxUV}) and  (\ref{eq:CWapproxIR}) are compared in Figure \ref{fig:CWplot}.

We should stress that in this analysis $B$ and $| \vec \phi|$ have been treated as background fields for the massive hypermultiplet. In the full theory, these fields are dynamical and one needs to extremize the full potential that includes, besides the Coleman-Weinberg contribution, the kinetic terms for these fields as well as the source term. In particular, extremizing with respect to $B$ yields a potential as a function of $|\vec \phi|$ and $\hat A_0$ only; at small values of the Coulomb branch field this prediction turns out to agree with the result from the magnetic theory, discussed below.

We conclude that a lattice of external vortices leads to a quantum instability --a runaway towards infinity-- along the Coulomb branch. In the magnetic description, this instability will arise from the condensation of scalar fields in the presence of a chemical potential. Anticipating the results from the mirror description, we point out that this instability will not gap the emergent Fermi surface, so in itself it is not an obstruction for our mechanism. 

\begin{figure}[h!]
\begin{center}
\includegraphics[width=0.45\textwidth]{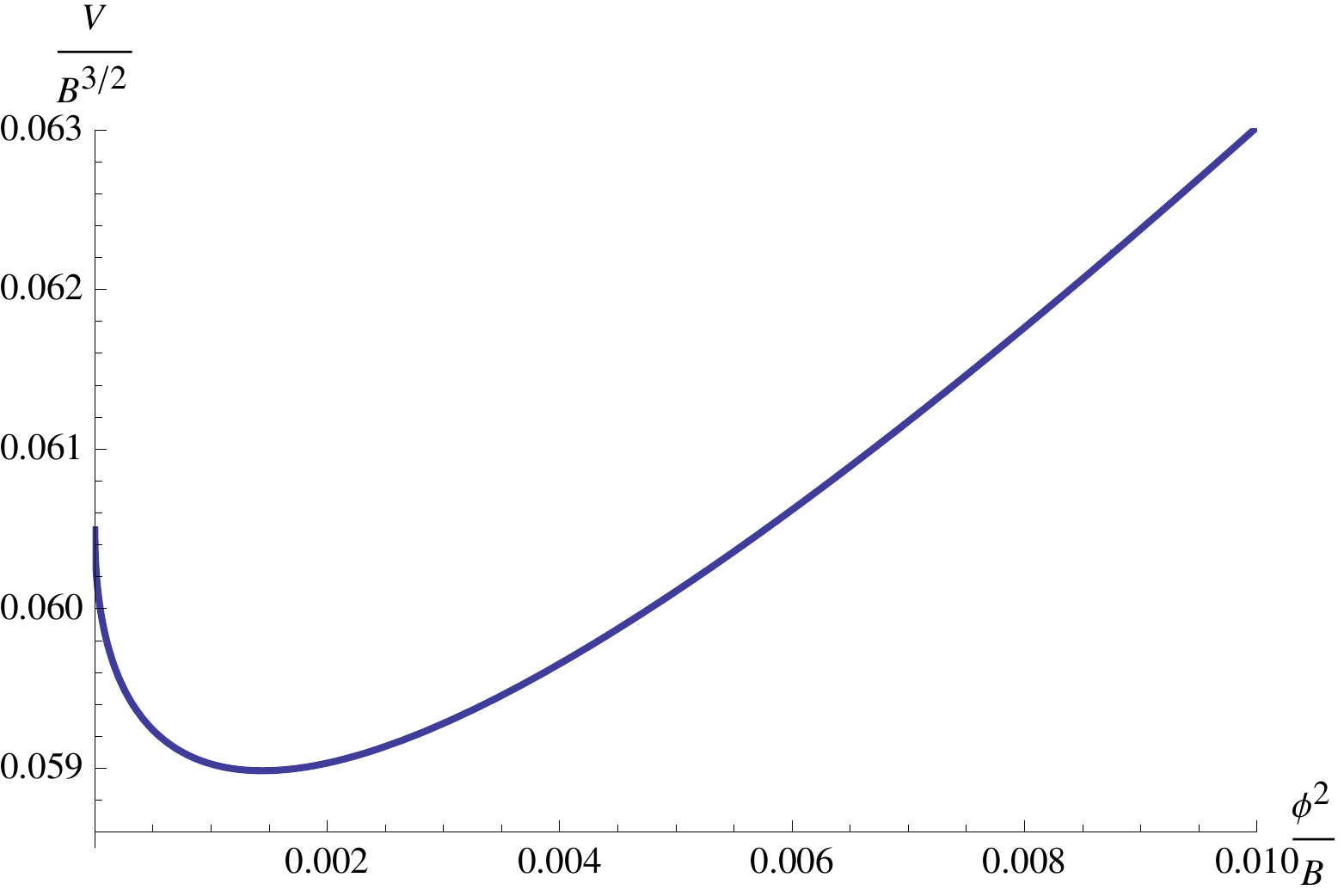}
\includegraphics[width=0.45\textwidth]{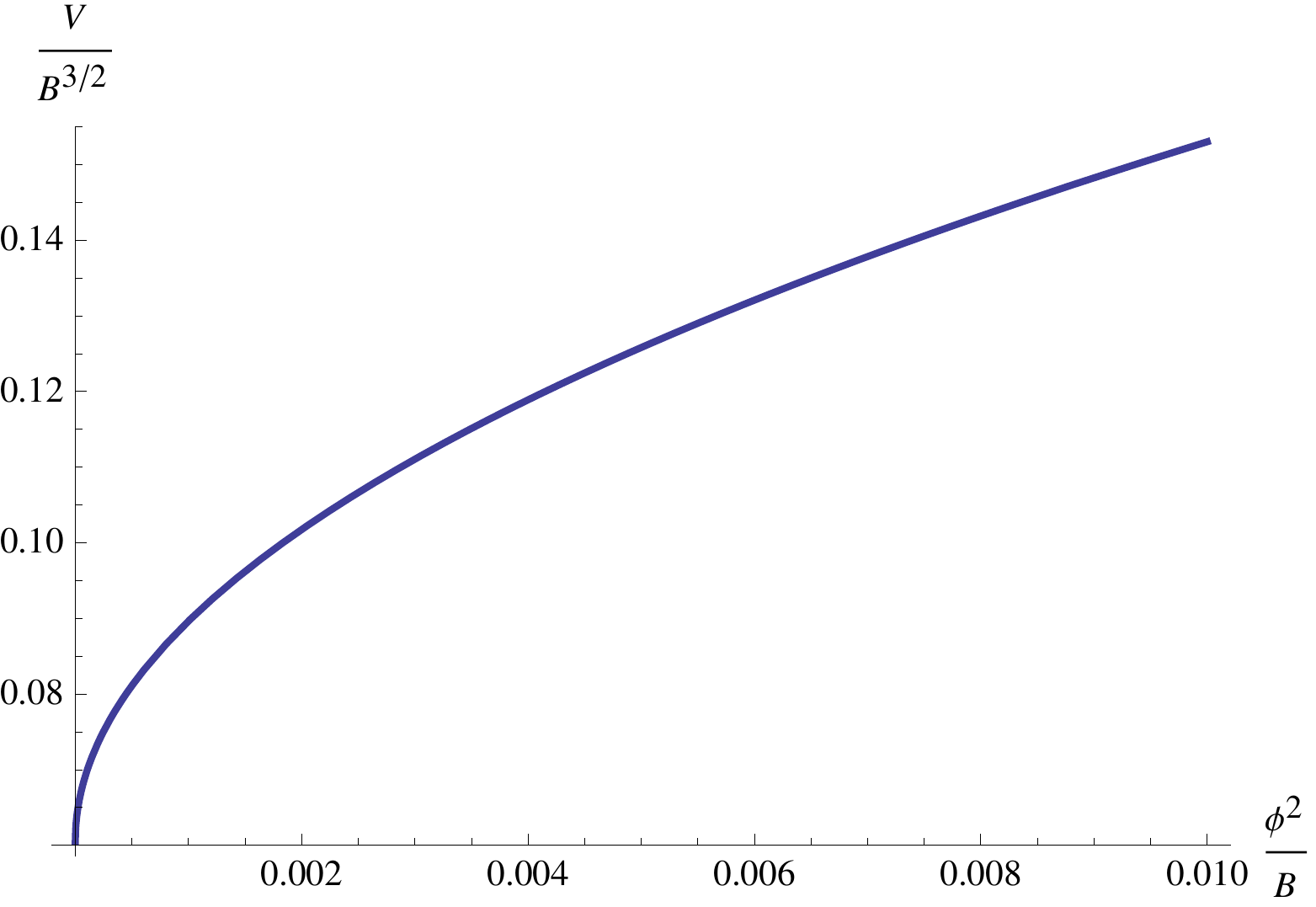}
\end{center}
\caption{\small{Coulomb-branch potential including the supersymmetry breaking deformation (\ref{eq:Vnonsusy}), for $\alpha=2$ on the left, and $\alpha=1$ on the right. The nonanalytic behavior near the origin  $| \vec \phi|^2=0$ becomes analytic in terms of the weakly coupled magnetic variables.}}\label{fig:liftedC}
\end{figure}

Nevertheless, we will be interested in stabilizing the runaway. A simple way to accomplish this is to add supersymmetry breaking terms that lift the Coulomb branch. As long as these deformations are much smaller than the gauge coupling, their effect can be analyzed very explicitly both in the electric and magnetic theory, using the mirror map derived before. In the electric theory, we can add a nonsupersymmetric term
\be\label{eq:Vnonsusy}
V \supset c_\alpha |\vec \phi|^\alpha\,,
\ee
which is allowed by the global symmetry group. This maps to $V \supset |v_i|^{2\alpha}$ in the magnetic description. Note that any $\alpha>0$ lifts the runaway (\ref{eq:CWapproxUV}). In particular, adding (\ref{eq:Vnonsusy}) to the one loop potential, we find that $\alpha=2$ (a soft supersymmetry breaking mass) gives a minimum away from the origin, while for $\alpha=1$ the minimum is at the origin for $c_\alpha\ge |B|/(4\pi)$.\footnote{We discuss $\alpha=1$ here even though it is not analytic in terms of the electric variables, because it becomes analytic in the magnetic dual variables, which are more relevant near the origin as they have a smooth K\"ahler potential.} These two possibilities are shown in Figure \ref{fig:liftedC}. We will find the same behavior using the magnetic description, thus confirming the duality with nonsupersymmetric deformations.

\subsection{Emergent Fermi surface in the magnetic description}

At low energies, the theory becomes strongly coupled and the appropriate description is in terms of the magnetic variables. In the presence of the $U(1)_J$ chemical potential the magnetic theory Lagrangian becomes
\be
L = |(\partial_0 - i \hat A_0) v_i|^2 - |\partial_j v_i|^2 + i \bar \psi_a \left[\gamma^0 (\partial_0 - i \hat A_0)+ \gamma^j \partial_j \right] \psi_a\,.
\ee
Since $v_i$ and $\psi_a$ are charged under $U(1)_J$ (see table (\ref{table:magnetic})), the chemical potential appears in the covariant derivatives for both fields. In the low energy limit they are approximately decoupled, and hence we obtain a superfluid $\langle v_i \rangle \neq 0$ coexisting with a Fermi surface.

In this way, $\mc N=4$ SQED provides an explicit and calculable example of an emergent Fermi surface. The microscopic theory has a nonzero chemical potential with no charged fermionic excitations (or, equivalently, a lattice of bosonic magnetic impurities), and in the IR there are composite charged fermions that organize into a Fermi liquid. The $U(1)_J$ charge in the UV is proportional to the volume of the Fermi surface in the IR. This gives an analytic realization of a Bose metal~\cite{bose}.

As it stands, the theory has an instability where the scalars condense, breaking $SU(2)_L \times U(1)_J \to U(1)$. This instability is the dual of the Coleman-Weinberg instability for the Coulomb branch in the presence of magnetic impurities. The latter is quantum-mechanical, while the former is seen in the tree level mirror description. Moreover, the condensation of the scalar fields does not lift the Fermi surface. In fact, the superfluid and the Fermi surface are distinguished by the global $SU(2)_L$ an $SU(2)_R$ symmetries.

Let us discuss how to stabilize the scalar fields. Following the discussion of the electric theory, we add a nonsupersymmetric potential for the $v_i$ consistent with the global symmetries which, combined with the contribution from the chemical potential, reads
\be
V = \t c_\alpha (|v_+|^2 + |v_-|^2)^\alpha - \hat A_0^2 (|v_+|^2 + |v_-|^2)\,.
\ee
The form of the nonsupersymmetric potential is obtained from (\ref{eq:Vnonsusy}) using the map between electric and magnetic variables.
The most relevant nonsupersymmetric deformations that are analytic in the magnetic variables and stabilize the runaway correspond to $\alpha=1,\,2$. For $\alpha=1$ and $\t c_\alpha> \hat A_0^2$, we find a stable minimum at $\langle v_i\rangle=0$, while for $\alpha=2$ there is a minimum away from the origin corresponding to a stable superfluid phase. This matches precisely the behavior that we found in the electric theory with supersymmetry breaking deformations (including the fact that for $\alpha=1$ the origin is stable for large enough $\t c_\alpha$), thus providing a very nontrivial check for our mapping of supersymmetry breaking deformations.

In summary, the possible long distance phases of $\mc N=4$ SQED doped with a uniform density of magnetic impurities (a $U(1)_J$ chemical potential) are
\begin{itemize}
\item[A)] a symmetric phase with an emergent Fermi surface described by a weakly coupled Fermi liquid, together with gapped bosons
\item[B)] an emergent Fermi surface coexisting with a stable superfluid phase, with symmetry breaking pattern $SU(2)_L\times SU(2)_R \times U(1)_J \to SU(2)_R \times U(1)$.
\end{itemize}
Note that in both cases the $U(1)_J$ charge of the microscopic theory is proportional to the volume of the Fermi surface at low energies. Phase A) is an example of emergent compressible quantum matter. It provides a strongly coupled but fully calculable example supporting the conjecture of~\cite
{Huijse:2011hp} that systems with a chemical potential for a global $U(1)$ symmetry, where the symmetry and translations are unbroken, have Fermi surfaces.

We note that this model provides a clean example of `hybridization' of the sort thought to occur in Kondo lattice models, where magnetic defects with tightly bound electrons contribute their electron count to the size of the Fermi surface at low energies.  Here, the magnetic defects are visible in the UV electric theory, while the IR magnetic theory sees a Fermi surface of itinerant fermions whose volume precisely captures the defect density.\footnote{By the index theorem for the Dirac operator in two dimensions, the number of fermionic zero modes localized on each magnetic impurity is given by the magnetic flux, which agrees with the charge density of the dual Fermi surface. See~\cite{Tong:2013iqa} for an analysis of more general spatially dependent impurities and their effects on vortices. }

\section{Fractionalization and emergent gauge fields in multiflavor SQED}\label{sec:N4gen}

Having understood the dynamics of SQED with one flavor in the presence of magnetic impurities, we now generalize the theory to include $N_f$ electrons. We will see that at low energies the electrons fractionalize and emergent gauge fields appear. Using mirror symmetry we will also argue that the theory of a Fermi surface coupled to a gauge field is dual to a model with external magnetic impurities. While both systems turn out to be strongly coupled in the IR, this new perspective may be helpful for understanding the dynamics of non-Fermi liquids with gapless bosons.

\subsection{Fractionalization and emergent gauge fields}\label{subsec:fractionalization}

Before proceeding to the explicit analysis, it is useful to make contact with some of the ideas of condensed matter physics that will be realized in our setup.  Pedagogical discussions of these
ideas can be found in e.g. \cite{WenLeeNagaosa,SSLee,Fradkin}.

Consider a lattice of spin 1/2 degrees of freedom coupled antiferromagnetically,
\be
H = J \sum_{\langle ij \rangle} \vec S_i \cdot \vec S_j + \ldots
\ee
The antiferromagnetic coupling $J>0$ tries to align the spins in antiparallel directions, and at low temperatures spins typically order and break the $SU(2)$ spin symmetry. However, there are systems where additional contributions tend to disorder the ground state, such as frustrated lattices, quantum fluctuations, etc. In order to describe these cases, the lattice spin operators are decomposed into  two-component spinors 
\be\label{eq:spinon}
\vec S_i = \sum_{\alpha \beta }f^\dag_{i\alpha} \vec \sigma_{\alpha \beta} f_{i \beta}
\ee
where $\vec \sigma$ are the Pauli matrices. 

The $f_i$ are called `spinons' because they carry spin but no charge and the decomposition of the $\vec S_i$ into the bilinear (\ref{eq:spinon}) is known as `fractionalization'. An important aspect of this representation is that it has a gauge redundancy $f_{i\alpha} \to e^{i \varphi_i} f_{i\alpha}$. The effective theory for $f_i$ then includes an emergent gauge field, with a kinetic term that can be generated quantum-mechanically. Such gauge fields lead to strong interactions between the spinons even in the absence of magnetic order, and are thought to be responsible for the formation of spin liquid states. 

Similar ideas appear in the context of non Fermi liquids. Propagating electrons $\psi_{i\alpha}$ (where $i$ is the lattice site and $\alpha$ is the spin index) may be fractionalized into degrees of freedom that carry the spin and the charge:
\be\label{eq:spinon-holon}
\psi_{i\alpha} = f_{i\alpha} b_i^\dag\,.
\ee
Here $b^\dag_i$ creates a hole at site $i$, while $f_{i\alpha}$ destroys a spin. The spinon carries the $SU(2)$ spin, while the `holon' $b_i$ carries the electromagnetic charge. As before, this description has a gauge redundancy $f_{i\alpha} \to e^{i \varphi_i} f_{i\alpha}$, $b_i \to e^{i \varphi_i} b_i$ and at low energies we expect a dynamical gauge field. Therefore, the effective theory will have a Fermi surface interacting with a gauge field. Understanding the dynamics of this strongly coupled system is an important open problem in condensed matter physics.

We note that fractionalization is somewhat counter to the intuition from relativistic gauge theories in four dimensions, where the strong dynamics tends to confine degrees of freedom at low energies. Here instead, degrees of freedom are deconfined at long distances. Nevertheless, we will now see that mirror symmetry leads precisely to relations like (\ref{eq:spinon}) and (\ref{eq:spinon-holon}) between the UV and IR fields.

\subsection{Mirror symmetry for SQED with $N_f$ flavors}\label{subsec:N4mirrorNf}

Mirror symmetry for $\mc N=4$ SQED with $N_f$ flavors was originally proposed in~\cite{Mirror}, and can be nicely derived from the $N_f=1$ case by taking $N_f$ powers of the superdeterminant formula of Kapustin and Strassler~\cite{Kapustin:1999ha}.

The electric description is given by $U(1)$ SQED with $N_f$ electron hypermultiplets $\mc Q_{Ii}= (Q_I,\,\t Q_I^*)$, $I=1,\,\ldots,\,N_f$. The global symmetries are
\be\label{eq:global-electric}
SU(2)_R \times SU(2)_L \times U(1)_J \times SU(N_f)
\ee
where the action of the R-symmetries $SU(2)_R \times SU(2)_L$ is the same as in \S \ref{sec:N4}, $U(1)_J$ is the topological symmetry dual to the gauge $U(1)$, and $SU(N_f)$ is the flavor symmetry for the $N_f$ hypermultiplets. The theory has a Coulomb branch of real dimension 4 which is parametrized by the dual photon $\gamma$ and the triplet of scalars $\vec \phi$ in the $\mc N=4$ vector multiplet. The Higgs branch has real dimension $4(N_f-1)$, and is given by the space $\mathbb C^{2N_f}$ of $ (Q_I,\,\t Q_I^*)$, modulo D- and F-term constraints,
\be
\sum_{I=1}^{N_f}(|Q_I|^2 - |\t Q_I|^2)=0\;,\;\sum_{I=1}^{N_f}\,Q_I \t Q_I=0\,.
\ee
The origin of moduli space flows in the IR to an interacting CFT, which is the focus of this work.
Also, semiclassically along the Coulomb branch there are vortex creation operators
\be
V_\pm \propto e^{\pm2\pi i \gamma /g^2}
\ee
where $\gamma$ is the dual photon introduced in \S \ref{sec:N4}.

\begin{figure}[h!]
\begin{center}
\includegraphics[width=0.5\textwidth]{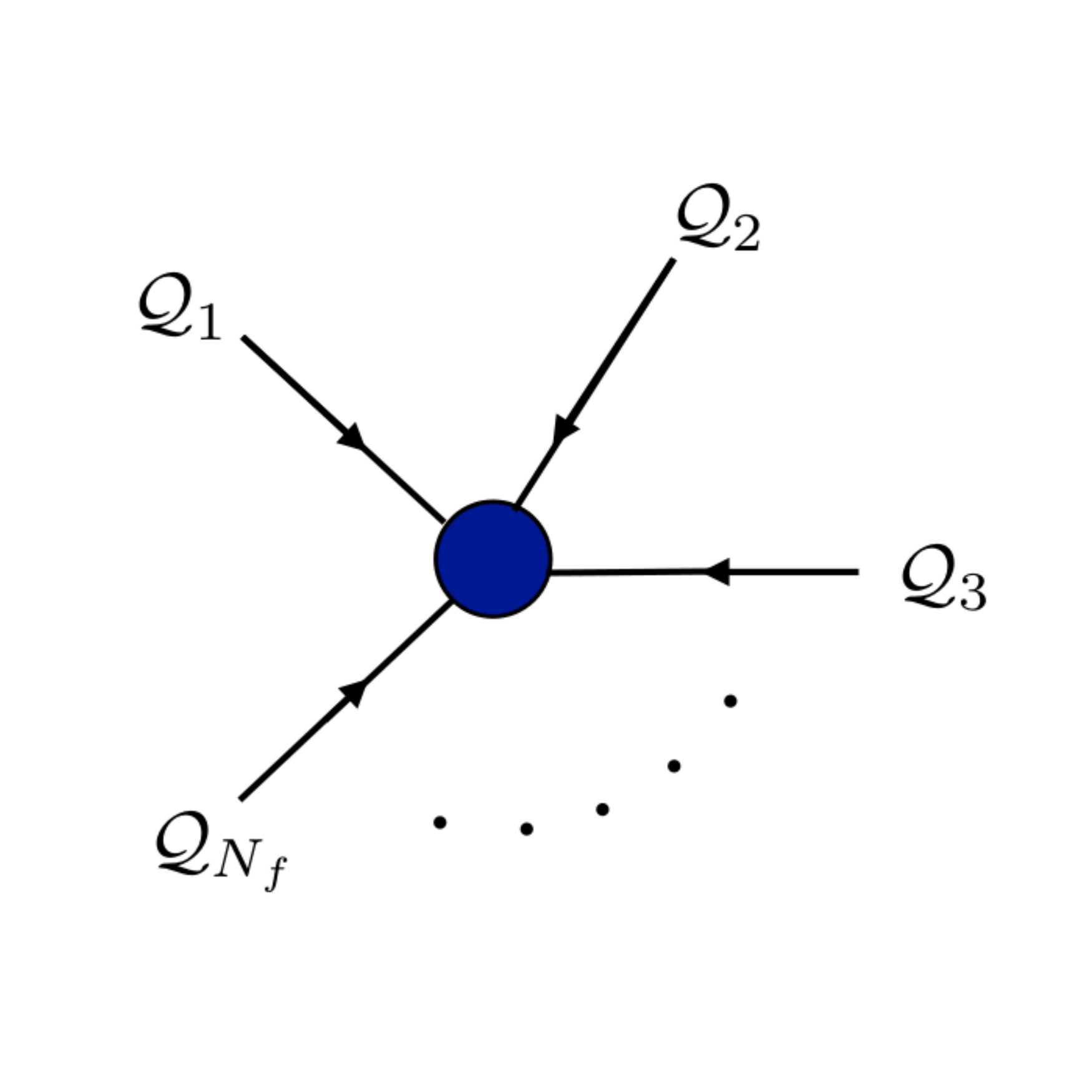}
\includegraphics[width=0.7\textwidth]{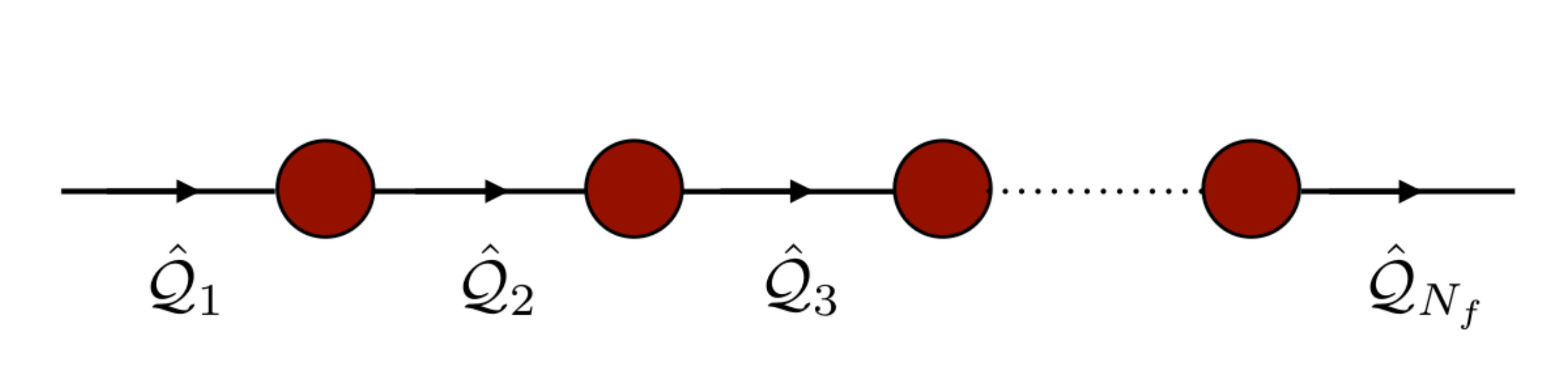}
\end{center}
\caption{\small{Quiver theories for mirror symmetry with $N_f$ flavors. On top, the electric $U(1)$ gauge theory with $N_f$ flavors $\mc Q$. On the bottom, magnetic dual with gauge group $U(1)^{N_f-1}$ and $N_f$ flavors $\mc{\hat Q}$. A node represents a $U(1)$ gauge factor, and an incoming arrow is a positively charged hypermultiplet.}}\label{fig:quiver}
\end{figure}

The mirror description is a quiver gauge theory with magnetic gauge group $U(1)^{N_f-1}$ and $N_f$ hypermultiplets $\mc{\hat Q}_{Ii}$. The $I$-th hypermultiplet is a bifundamental with charge $+1$ under the $I$-th $U(1)$ factor and charge $-1$ under $U(1)_{I-1}$. Here $U(1)_0$ and $U(1)_{N_f}$ are not part of the gauge group, so that the first and last hypermultiplets are fundamentals (instead of bifundamentals) of $U(1)_1$ and $U(1)_{N_f-1}$ respectively.\footnote{Using the result of~\cite{Kapustin:1999ha}, this quiver is obtained by starting from a $U(1)^{N_f}$ theory with $N_f$ flavors where the $I$-th flavor is charged under the $I$-th $U(1)$ factor, and then setting to zero the total sum of the $U(1)$ generators.}  The dual electric and magnetic quiver gauge theories are shown in Figure \ref{fig:quiver}.

The global symmetries of the magnetic theory at the classical level are
\be\label{eq:global-magnetic}
SU(2)_R \times SU(2)_L \times U(1)_J^{N_f-1} \times U(1)\,.
\ee
Besides the R-symmetries, there are now $N_f-1$ topological global symmetries from dualizing the $N_f-1$ $U(1)$ factors, and an additional global $U(1)$ under which all the $\mc{\hat Q}$ have charge $1/N_f$ (the normalization is set by the mapping of vortices, explained below). This model has a Coulomb branch of real dimension $4(N_f-1)$, and a Higgs branch of dimension 4.

The mapping of global symmetries under the duality is more nontrivial than in the $N_f=1$ case. The electric theory has a nonabelian $SU(N_f)$ flavor symmetry. Its Cartan subgroup $U(1)^{N_f-1}$ maps to the topological $U(1)_J^{N_f-1}$. This means that at strong coupling, the magnetic $U(1)_J^{N_f-1}$ should be enhanced to a full $SU(N_f)$ -- the details of this, which depend on properties of vortex operators, are not completely understood yet. On the other hand, the topological $U(1)_J$ of the electric theory maps to the global $U(1)$ of the magnetic theory under which all the $\hat{\mc Q}$ have the same charge $1/N_f$. The charge assignment is a consequence of the vortex mapping,  which can be read off from the hyperkhaler metric of the magnetic theory. The unit-charge vortex operators $V_\pm$ (which semiclassically behave like $V_\pm \sim e^{\pm 2\pi i \gamma/g^2}$) are identified with products of elementary magnetic flavors,
\be
V_+ \,\leftrightarrow\, \hat Q_1 \ldots \hat Q_{N_f}\;,\;V_- \,\leftrightarrow\,  \hat{\tilde Q}_1 \ldots \hat{\tilde Q}_{N_f}
\ee
Note that for $N_f=1$, this reproduces what we saw in \S \ref{sec:N4} -- the vortices of the electric theory become elementary fields in the IR. In contrast, for $N_f>1$ the integer-charged vortices of the electric theory split into vortices $\hat{\mc Q}$ of fractional charge. However, the fractional charge cannot be observed in gauge invariant operators.

Under mirror symmetry, the Higgs and Coulomb branch are interchanged. More specifically, the triplet of Coulomb branch scalars of the electric theory maps to the Higgs branch coordinates of the magnetic theory,
\be\label{eq:CHfrac}
\vec \phi\,\leftrightarrow\, \sum_I\,\hat q^\dag_{Ii} \vec \sigma_{ij} \,\hat q_{Ij}\,,
\ee
where $\hat q$ is the scalar component of the hypermultiplet $\hat{\mc Q}$. The precise numerical coefficient in this map, which we will not need, can be calculated from the Taub-NUT metric generated by integrating out $N_f$ flavors in the electric theory.

At this point we can compare with the discussion of \S \ref{subsec:fractionalization}. We see that (\ref{eq:CHfrac}) has precisely the structure (\ref{eq:spinon}) for the fractionalization of lattice spins into spinons. Moreover, in the sum over the magnetic flavors there are $N_f$ phase ambiguities. Of these, $N_f-1$ are the emergent gauge fields of the magnetic quiver theory, while the sum of all the phases is the $U(1)$ global symmetry in (\ref{eq:global-magnetic}). The appearance of this global symmetry is special to three dimensions where, as we discussed before, a gauge symmetry gives rise to a conserved current; also, in this case the dual photon appears as the overall (dynamical) phase of the magnetic flavors.

In this way, mirror symmetry provides an analytically controlled realization of fractionalization and emergent gauge fields. The separation into spinons (\ref{eq:CHfrac}) is determined (and protected) by $\mc N=4$ SUSY. There is also a version of the separation of an electron into a spinon-holon pair, Eq.~(\ref{eq:spinon-holon}). Indeed, the SUSY partner of (\ref{eq:CHfrac}) implies that the electric theory gaugino corresponds to a product of scalar and fermion flavors in the magnetic theory: $ \lambda_{ia}\,\leftrightarrow\, \sum_I \hat q^\dag_{Ii} \,\hat\psi_{Ia}+c.c.$, where $\hat\psi_{Ia}$ is the superpartner of $\hat q_{Ii}$. We thus find spin and charge separation into $\hat q$ and $\hat \psi$; instead of the abelian electromagnetic charge here we have nonabelian $SU(2)$ symmetries.

\subsection{Duality for a Fermi surface coupled to a gauge field}\label{subsec:Nfmagnetic}

An important aspect of the duality is that the topological $U(1)_J$ of the electric theory is mapped to the global $U(1)$ in the magnetic theory, under which all the magnetic flavors have charge $1/N_f$. This generalizes the mapping studied in \S \ref{sec:N4}. A background vector multiplet $\hat{\mc V}$ for $U(1)_J$ then appears in the electric theory via the BF interaction (\ref{eq:LBF}), while in the magnetic theory it enters through the covariant derivatives for the $\hat{\mc Q}$ flavors. In fact, the situation is now more symmetric between the electric and magnetic theory, because the topological $U(1)_J^{N_f-1}$ symmetries of the magnetic theory, which enter the action through BF interactions, map to flavor symmetries $U(1)^{N_f-1} \subset SU(N_f)$ in the electric theory. This will become even more explicit in the generalization discussed in \S \ref{subsec:Dbrane}.

Let us now add external impurities, focusing for simplicity on the case $N_f=2$. The electric and magnetic theories are self-dual (both are $U(1)$ gauge theories with two flavors), though the mapping of global symmetries is still nontrivial. In particular, let us add a lattice of magnetic impurities to the electric description which, at distances larger than the lattice spacing, amounts to a $U(1)_J$ chemical potential. The dual is a $U(1)$ gauge theory with a chemical potential for the two flavors. 

The weakly coupled dynamics of the electric theory is a straightforward generalization of the $N_f=1$ discussion. The lattice of magnetic impurities sources a magnetic field; the fermionic components of ${\mc Q}_I$ have zero modes localized on each impurity, with degeneracy proportional to the magnetic charge. The scalar partners are gapped. Averaging the magnetic field over distances larger than the lattice spacing gives two copies of Landau levels (\ref{eq:LL}). The lowest Landau level of the fermions has $E=0$ and a ground state degeneracy that agrees with the number of zero modes localized at each impurity times the area. The scalars Landau levels are gapped. The Coulomb branch fields are classically massless but acquire a one-loop Coleman-Weinberg instability. As we discussed in \S \ref{subsec:N4el1}, the instability is lifted by adding supersymmetry breaking deformations $V \supset |\vec \phi|^\alpha$. Therefore, the gapless degrees of freedom are a gauge field interacting with fermion zero modes $\psi_{Ii}$ localized at the magnetic impurities.

Because of the emergent gauge field, the dynamics in the magnetic theory for $N_f > 1$ is much more interesting than the $N_f=1$ case.  At tree level we have a chemical potential for the global $U(1)$ under which the two hypermultiplets $(\hat{\mc Q}_1,\,\hat{\mc Q}_2)$ have the same charge. The scalar components are thus tachyonic at tree level, and the instability is lifted by a nonsupersymmetric deformation that is dual to the one introduced in the electric theory. There are additional massless scalars from the vector multiplet, which are lifted by quantum corrections --this is dual to the tree level statement that the Higgs branch of the electric theory is gapped by the magnetic field. In summary, the magnetic theory gives rise to a Fermi surface for the fermions $(\hat \psi_I, \hat{\tilde \psi}_I)$, $I=1,2$, interacting with a $U(1)$ gauge field.

In this way, we obtain a rather surprising duality between the theory of a Fermi surface for two Dirac fermions interacting with a gauge field, and a gauge theory with external magnetic impurities and two charged Dirac fermions.
This generalizes to $N_f>2$ by having extra gauge fields coupled to the Fermi surface in the magnetic side, and additional electrons interacting with the lattice of impurities in the electric description. The theory of a gauge field interacting with a Fermi surface plays an important role in recents attempts to understand non-Fermi liquids and high temperature superconductors (for early references see
\cite{Polchinski:1993ii,wen,millis}), so its realization within SQED and its dual description are results that could have more general relevance. Both models are strongly coupled, but in a very different way. It would be very interesting to analyze in detail the dynamics of the model with magnetic impurities and its large $N_f$ limit.

\subsection{D-brane interpretation and further generalizations}\label{subsec:Dbrane}

Let us end our analysis by reviewing the D-brane realization of the $\mc N=4$ theories and mirror symmetry~\cite{Hanany:1996ie,deBoer:1996mp,deBoer:1996ck}, which provides a geometric interpretation for some of the phenomena that we have encountered.

\begin{figure}[h!]
\begin{center}
\includegraphics[width=0.8\textwidth]{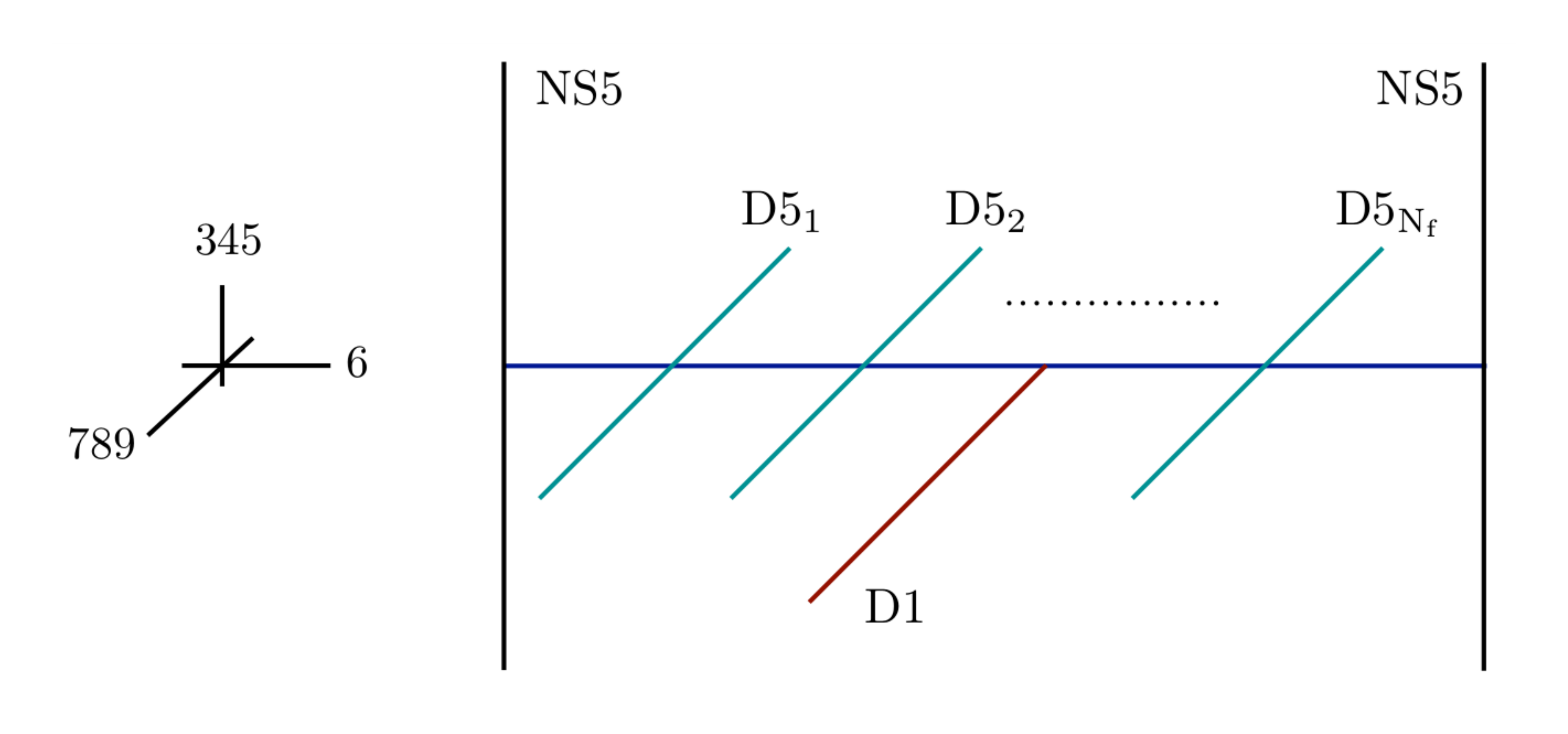}
\end{center}
\caption{\small{D-brane model for $U(1)$ SQED with $N_f$ flavors, with one magnetic impurity (D1 brane) inserted. The common space-time directions $(012)$ are not shown.}}\label{fig:brane1}
\end{figure}

Consider the electric theory with $N_f$ flavors in type IIB string theory. The three dimensional $U(1)$ vector multiplet is given by a single D3 brane extended along the spacetime directions $(0126)$ and ending between two NS5 branes that are extended along $(012345)$ and are at different points in the 6-th direction. The flavors come from $N_f$ D5 branes along $(012789)$. A lattice of magnetic impurities is given by inserting D1 branes\footnote{These are magnetic monopoles from the point of view of the D3-brane worldvolume.} of infinite length, extended along $(03)$ and located at points on the field theory spatial directions $(12)$. The D-brane construction is shown in Figure \ref{fig:brane1}.

\begin{figure}[h!]
\begin{center}
\includegraphics[width=0.8\textwidth]{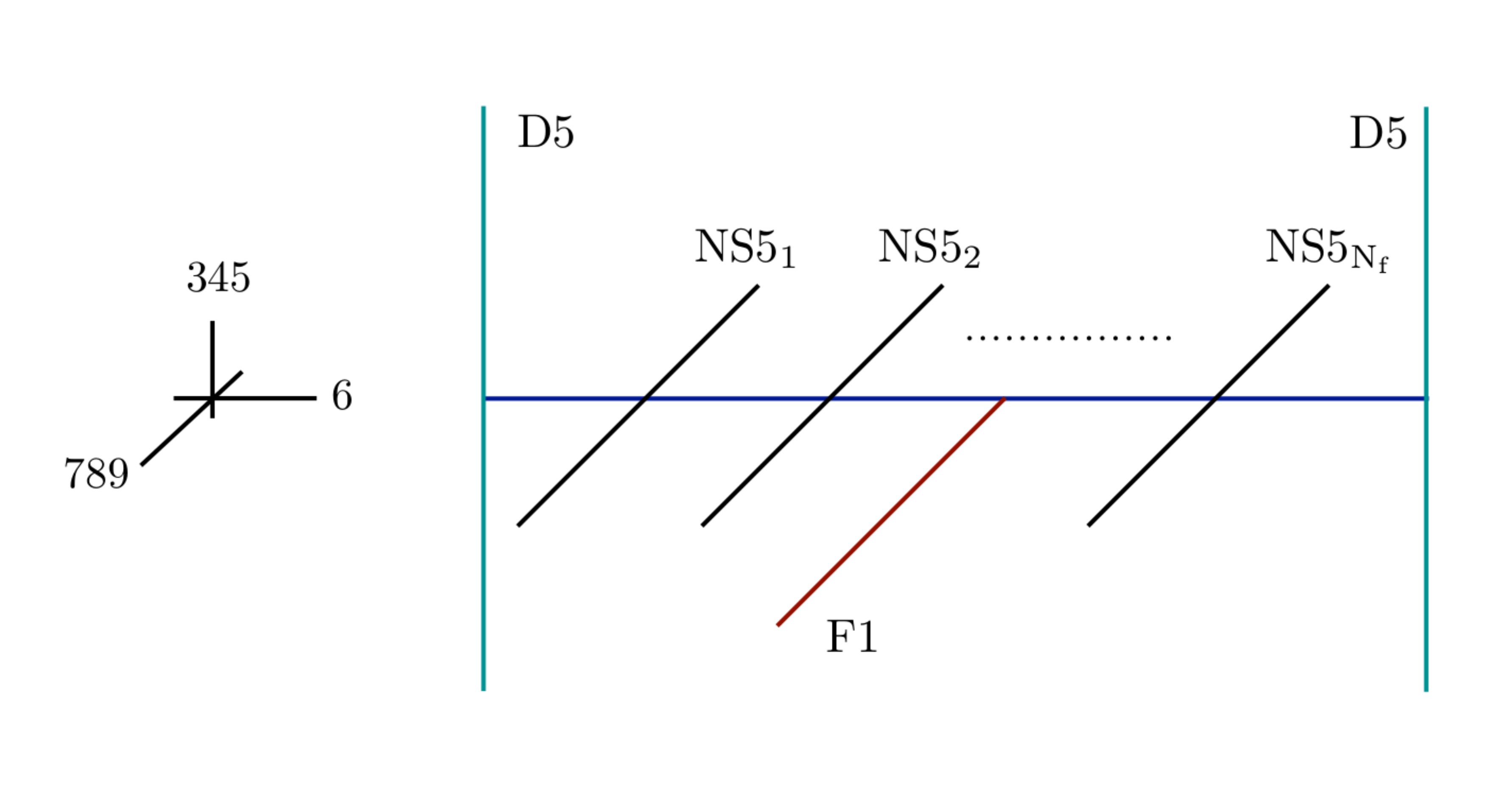}
\end{center}
\caption{\small{D-brane model for the magnetic description, a $U(1)^{N_f-1}$ SQED with $N_f$ flavors. Also shown is an insertion of an electric impurity (F1 string).}}\label{fig:brane2}
\end{figure}

The dual description is obtained by applying the S-duality of type IIB string theory to this brane configuration. Recalling that the D3 is self-dual, and that S-duality exchanges D1 and F1, and D5 and NS5, respectively, we obtain the D-brane system of Figure \ref{fig:brane2}. The matter content of the magnetic quiver is now clear: each D3 segment between two adjacent NS5 branes gives rise to a $U(1)$ multiplet, so the gauge group is $U(1)^{N_f-1}$. Open strings stretched between adjacent D3 segments are the bifundamental fields. The first and last D3 segments, which extend between a D5 and NS5, have no worldvolume degrees of freedom; the open strings that stretch across the first/second and $(N_f-1)$/$N_f$ D3 segments are associated to $\hat{\mc Q}_1$ and $\hat{\mc Q}_{N_f}$, respectively. Furthermore, the impurity has become an F1 string, which couples electrically to the charged fields, hence giving rise to a chemical potential.

One aspect to note in this construction is that the external sources represented by the D1s or F1s preserve half of the supercharges. The effect of these BPS sources was studied in~\cite{Hook:2013yda}, where it was shown how some of the supersymmetries are preserved by turning on `superpartners' of the chemical potential or magnetic field. Here, instead, we have added nonsupersymmetric impurities, setting these superpartners to zero. It would be interesting if such deformations admit a D-brane realization, either by rotating the branes through different angles or by turning on additional worldvolume parameters.

Mirror symmetry can be extended to more general abelian theories~\cite{deBoer:1996ck}. For instance, in the previous D-brane realization of the electric theory we can consider $r+1$ NS5 branes (instead of just two). This makes the duality between electric and magnetic descriptions more symmetric: the electric theory has $r+1$ NS5 branes and $N_f$ D5 branes, while the magnetic dual has $r+1$ D5 branes and $N_f$ NS5 branes. As a result, mirror symmetry relates
\begin{itemize}
\item an $\mc N=4$ $U(1)^r$ gauge theory with $N_f$ hypermultiplets $\mc Q_I$
\item an $\mc N=4$ $U(1)^{N_f-r}$ gauge theory with $N_f$ hypermultiplets $\hat{\mc Q}_I$\,.
\end{itemize}
The structure of the quivers can be easily deduced from the brane construction, as in the $r=1$ case that we just discussed. The global (non R-symmetries) of these theories include $U(1)^{N_f-r}_F \times U(1)^r_J$ in the electric theory, and $U(1)^r_F \times U(1)^{N_f-r}_J$ in the magnetic dual. Mirror symmetry exchanges the flavor and topological symmetries of both sides.

We can now add impurities to this duality in different ways. For instance, a lattice of magnetic impurities for one of the $U(1)$ gauge factors in the electric theory will result in a chemical potential for one of the flavor symmetries in the mirror. This generalizes the duality of \S \ref{subsec:Nfmagnetic} for the Fermi surface coupled to a gauge field by adding more relativistic degrees of freedom (e.g. extra gauge fields and bifundamentals) that are coupled to the Fermi surface. It would be interesting to analyze the dynamics of these models.

\section{Emergent non-Fermi liquid in $\mc N=2$ SQED}\label{sec:N2}

So far we have considered $\mc N=4$ theories with impurities and the dynamics predicted by mirror symmetry. For $N_f=1$ we obtained a weakly coupled Fermi liquid of vortices, while larger $N_f$ led to emergent gauge fields. Now we will study theories with half of the supersymmetries ($\mc N=2$), which lead to novel effects --such as SUSY Wilson-Fisher fixed points-- not observed in their $\mc N=4$ partners. In particular, we focus on the
 $\mc N=2$ theory with $U(1)$ gauge group and $N_f=1$, for which already both the electric and magnetic description are strongly coupled in the IR. The addition of magnetic impurities will then lead to an interacting Fermi surface of vortices.

\subsection{Mirror symmetry in $\mc N=2$ SQED}\label{subsec:N2mirror}

Let us review the duality for the $\mc N=2$ theory~\cite{deBoer:1997ka,deBoer:1997kr,Aharony:1997bx}. 

The electric theory  is obtained from $\mc N=4$ SQED by lifting the Coulomb branch chiral superfield $\Phi$. This may be accomplished by adding a new chiral superfield $S$ to the $\mc N=4$ theory, with superpotential
\be
W = S \Phi\,.
\ee
This lifts $\Phi$ and $S$ and breaks half of the supersymmetries, leaving $\mc N=2$ QED with $U(1)$ gauge group and one flavor $(Q, \t Q)$. Unlike the previous section, it is now convenient to work with $\mc N=2$ superfields, in terms of which the Lagrangian reads
\be
L_\text{el}= \int d^4 \theta\,\left( Q^\dag e^{2V} Q + \t Q^\dag e^{-2V} \t Q\right)+ \frac{1}{2g^2}\,\int d^2 \theta\,W_\alpha^2+c.c.+ L_{BF}\,,
\ee
where $L_{BF}$ includes the external sources (\ref{eq:LBF}) restricted to the $\mc N=2$ vector multiplet.

There are various important consequences of this deformation. First, the F-term conditions from $W = Q \Phi \t Q$ are now absent so, unlike the $\mc N=4$ theory, there is a nontrivial Higgs branch parametrized by the meson invariant
\be
M = Q \t Q\,.
\ee
Furthermore, in the $\mc N=2$ theory, the nonabelian global symmetries are broken down to $U(1)_R \times U(1)_A$, where $U(1)_R$ is the R-symmetry group (under which $Q$ and $\t Q$ have charge 0), and $U(1)_A$ is the sum of the unbroken $U(1)$ subgroups of $SU(2)_L \times SU(2)_R$  (under which $Q$ and $\t Q$ have charge 1). Another new effect is that due to a one loop Chern-Simons anomaly, the dual photon also acquires $U(1)_A$ and $U(1)_R$ charges~\cite{Aharony:1997bx}. The global symmetries including the topological $U(1)_J$ are then
\begin{center}
\be\label{table:N2globale}
\begin{tabular}{c|ccc}
&$U(1)_R$&$U(1)_A$&$U(1)_J$\\
\hline
&&&\\[-12pt]
$e^{\pm 2\pi (\sigma+i \gamma)/g^2}$  &1& -1 & $\pm 1$  \\
&&&\\[-12pt]
$\lambda$  & 1& 0 &  0 \\
&&&\\[-12pt]
$q$  & 0& 1 & 0  \\
&&&\\[-12pt]
$\psi_q$  & -1& 1 & 0 \\
&&&\\[-12pt]
$\t q$  & 0&1 & 0  \\
&&&\\[-12pt]
$\psi_{\t q}$  & -1& 1 & 0 
\end{tabular}
\ee
\end{center}

Classically, the natural holomorphic coordinate for the Coulomb branch is
\be
\Sigma = \sigma+ i \gamma\,.
\ee
As in the $\mc N=4$ theory, the Coulomb branch receives quantum-mechanical corrections, splitting into two regions $V_\pm$. Far along the Coulomb branch, $v_\pm \sim e^{\pm 2\pi i \Sigma/g^2}$, which are cylinders of radius $g$ (recalling the periodicity $\gamma \to \gamma+g^2$). Near the origin, where the Coulomb and Higgs branch intersect, $v_\pm$ shrink to zero size. Therefore, the moduli space near the origin is topologically the intersection of three cones $M$ and $V_\pm$.

The magnetic theory is the description of such a low energy region, in terms of the gauge invariant fields $M$ and $V_\pm$, subject to the superpotential
\be
W = \sqrt{2}\,h M V_+ V_-\,,
\ee
which is the most relevant interaction consistent with the symmetries that correctly reproduces the previous topological structure. The global symmetries follow from (\ref{table:N2globale}), with the identification $M = Q \t Q$ and $v_\pm \sim e^{\pm 2\pi i \Sigma/g^2}$:
\begin{center}
\be\label{table:N2globalm}
\begin{tabular}{c|ccc}
&$U(1)_R$&$U(1)_A$&$U(1)_J$\\
\hline
&&&\\[-12pt]
$V_\pm$  &1& -1 & $\pm 1$  \\
&&&\\[-12pt]
$M$  & 0& 2 & 0
\end{tabular}
\ee
\end{center}
We can also include external sources $\hat V$ for $U(1)_J$. In this case, 
 the Lagrangian  is
\be
L_\text{mag} =  \int d^4 \theta\,\left( V_+^\dag e^{2 \hat V} V_+ + V_-^\dag e^{-2 \hat V}  V_-+ M^\dag M\right)+ \int d^2 \theta\,\sqrt{2}h M V_+ V_-+c.c.
\ee

The magnetic theory has a relevant coupling $h$ (with dimension $[h^2]=1$), which determines the strength of quartic and Yukawa interactions. It flows to an interacting fixed point that is a supersymmetric generalization of the Wilson-Fisher fixed point in three dimensions. Mirror symmetry then relates $\mc N=2$ SQED at strong coupling $E \ll g^2$ to the supersymmetric Wilson-Fisher theory in the strongly coupled regime $E \ll h^2$. Unlike the $\mc N=4$ case, this is a strong-strong duality. In the $\mc N=4$ duality there was an exact map between the electric and magnetic variables, given by the Taub-NUT sigma model. However, for $\mc N=2$ mirror symmetry, we do not have such an explicit dictionary of the duality, except for the meson $M= Q \t Q$. The strong dynamics of the magnetic theory will have important consequences once we add the lattice of magnetic impurities. In particular, let us note for future use that the fermions $\psi_{\pm}$ arise as composites of the electric degrees of freedom,
\be
\psi_{\pm} = f(\sigma) e^{\pm2\pi i \gamma/g^2} \lambda\,,
\ee
where $\lambda$ is the SQED gaugino, and $f(\sigma)$ is some unknown function which, unlike the Taub-NUT case, receives corrections to all orders in perturbation theory. The dependence on the dual photon is fixed by global symmetries.

\subsection{Dynamics in the presence of magnetic impurities}\label{subsec:N2impurities}

We are now ready to add the lattice of external magnetic impurities to $\mc N=2$ SQED. As in the previous case, we work with a uniform distribution of impurities and at distances much larger than the lattice spacing, so that their effect is equivalent to a uniform $\hat A_0$ -- a chemical potential for the $U(1)_J$ symmetry.

The dynamics at weak coupling is very similar to that of \S \ref{sec:N4}, the main difference being the absence of $\Phi$ and of the superpotential coupling $W = \sqrt{2} Q \Phi \t Q$. At the classical level, the source $\hat A_0$ generates a constant magnetic field $B = \frac{g^2}{2\pi} \hat A_0$. The charged scalars and fermions are then in Landau levels with spectrum (\ref{eq:LL}). They give rise to a one-loop instability for the Coulomb branch (\ref{eq:CWfinal}), which is the same as in the $\mc N=4$ theory with the replacement $|\vec \phi|= |\sigma|$. This runaway towards large $\sigma$ can be avoided by adding a nonsupersymmetric potential for $\sigma$,
\be\label{eq:VnonsusyelN2}
V \supset |\sigma|^\beta\;,\;\beta>0\,.
\ee

Let us now discuss the dynamics from the point of view of the magnetic variables $V_\pm$ and $M$. At weak coupling, the dynamics is controlled by the Lagrangian
\bea
L_\text{mag}&=& |(\partial_0-  i \hat A_0)v_+|^2- |\partial_i v_+|^2 + |(\partial_0  +i \hat A_0)v_-|^2- |\partial_i v_-|^2+ i \bar \psi_+ \left[\gamma^0(\partial_0 - i \hat A_0)+ \gamma^i \partial_i \right] \psi_+\nonumber\\
&+& i \bar \psi_- \left[\gamma^0(\partial_0 + i \hat A_0)+ \gamma^i \partial_i \right] \psi_-+|\partial_\mu M|^2+ i \bar \psi_M \not \! \partial \psi_M - V_\text{mag}
\eea
where the potential
\be\label{eq:Vmag}
V_\text{mag} =2 |h|^2(|v_+v_-|^2 + |M v_+|^2+|M v_-|^2)+ \sqrt{2} h (M \psi_+ \psi_-+ v_+ \psi_M \psi_- + v_- \psi_M \psi_+)+c.c.
\ee
The theory suffers from a classical instability due to the chemical potential for the charged scalars. For instance, the quartic potential in (\ref{eq:Vmag}) vanishes along $v_-=M=0$, while the tachyonic contribution from the chemical potential lowers the energy by increasing $v_+$. This is the counterpart of the quantum instability in the electric theory.

The simplest way to lift the tachyon is by adding a nonsupersymmetric term
\be\label{eq:VnonsusymagN2}
V \supset c_\alpha (|v_+|^2 + |v_-|^2)^\alpha\,.
\ee
For $\alpha=1$ and $c_\alpha> \hat A_0^2$, the charged scalars are stabilized at the origin. It is also possible to have a vacuum away from the origin. In this case, the Yukawa interactions mass-up $\psi_M$ and one linear combination of the charged fermions $\langle v_+ \rangle \psi_- + \langle v_- \rangle \psi_+$. It is also important to note that the mapping between nonsupersymmetric deformations (\ref{eq:VnonsusyelN2}) and (\ref{eq:VnonsusymagN2}) is not known explicitly, unlike the $\mc N=4$ case. Nevertheless, all that we need for our purpose is that such a mapping exists, which is guaranteed by the matching of moduli spaces of the electric and magnetic theory.

To summarize, the possible phases of the magnetic theory at tree level  are
\begin{itemize}
\item[A)] a vortex Fermi surface interacting with a relativistic boson $M$ and fermion $\psi_M$;
\item[B)] a superfluid phase where some of the charged scalars condense, together with a massless fermion which, depending on the scalar condensate, could have a nonvanishing Fermi surface.
\end{itemize}

\subsection{One-loop analysis of the magnetic theory}\label{subsec:oneloop}

In what follows we will focus on the phase A) above, which arises when the scalars $v_\pm$ are stabilized at the origin. At tree level, the magnetic theory describes a Fermi surface for the fermionic vortices $\psi_\pm$, together with the meson chiral superfield $(M, \psi_M)$. We will describe what happens at one loop, and then present some guesses about the dynamics at strong coupling. 
The one-loop analysis becomes uncontrolled (even if the Yukawa coupling is perturbatively small) in
certain kinematic regimes, a fact familiar from many earlier studies of the theory of gapless bosons
coupled to a Fermi surface.  Nevertheless, it provides useful intuition, and is controlled in some
energy range (when the Yukawa coupling is sufficiently small).
It could be interesting to analyze a controlled version of this theory (for instance using large $N$), a point which we hope to address in the future.

Neutrality of the $U(1)_A$ charge requires
\be
\bar \psi_+ \gamma^0 \psi_+ + \bar \psi_- \gamma^0 \psi_- =0
\ee
so that the number density of $\psi_+$ and $\psi_-$ have the same magnitude and opposite sign, and the total $U(1)_J$ charge is proportional to twice the volume of each of the Fermi surfaces,
\be
Q = 2 \int d^2x\,\bar \psi_+ \gamma^0 \psi_+\,.
\ee
Let us now discuss the interactions between these degrees of freedom. The gapless boson $M$ couples to $\psi_\pm$ via a Yukawa interaction. The 3d Dirac fermion $\psi_M$ couples to $\psi_\pm$ via irrelevant quartic interactions, obtained by integrating out the massive $v_\pm$. Therefore, the  tree level potential for the light fields is
\be
V =\sqrt{2} h\,M \psi_+ \psi_- + c.c. - \frac{2|h|^2}{m^2}|\psi_M|^2 (|\psi_+|^2+ |\psi_-|^2)\,.
\ee

We now proceed to compute the one loop self-energies for the light fields.
The propagator for a field $\phi$ is denoted by  $G_\phi(p) = \langle \phi(p) \phi^*(p) \rangle$, and its one loop self-energy by $\Sigma_\phi$, so that
\be
G_\phi^{(1)}(p) = G_\phi^{(0)}(p)- G_\phi^{(0)}(p)\,\Sigma_\phi(p)  G_\phi^{(0)}(p) + \ldots = \frac{1}{ G_\phi^{(0)}(p)^{-1}+\Sigma_\phi(p)}\,.
\ee
Here $G^{(0)}$ and $G^{(1)}$ are the tree level and one loop Green's functions, respectively.

\begin{figure}[h!]
\begin{center}
\includegraphics[width=0.5\textwidth]{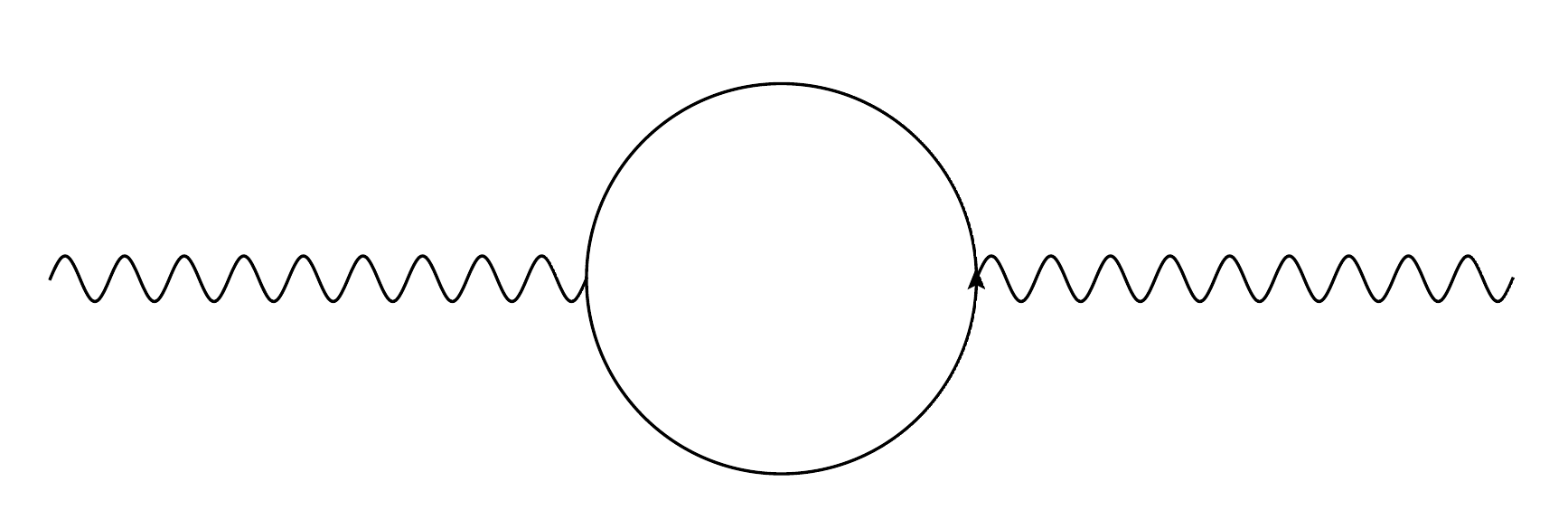}
\end{center}
\caption{\small{One-loop self-energy diagram for the bosonic field $M$.}}\label{fig:bosonself}
\end{figure}

For the boson $M$, evaluating the diagram in Figure \ref{fig:bosonself}, we find
\be
\Sigma_M(p)=- |h|^2 \int \frac{d^3q}{(2\pi)^3}\,G_{\psi_+}^{(0)}(q)G_{\psi_-}^{(0)}(-q-p)= |h|^2  \int \frac{d^3q}{(2\pi)^3}\,\frac{i}{\slashed q+ \hat{\slashed A}}\,\frac{i}{\slashed p+\slashed q+ \hat{\slashed A}}\,,
\ee
where $\hat A_\mu=(\mu,0,0)$ and the $i \epsilon$ prescription, not shown here, amounts to replacing $\slashed q \to \slashed q + i \epsilon$. The dependence on the chemical potential can be absorbed into a redefinition of the loop momentum, so the boson self-energy is the same as in the relativistic case.
Evaluating this integral one obtains\footnote{See~\cite{Peskin:1995ev} for the required formulas in $d$ dimensions and for the conventions which we follow here.}
\be\label{eq:GM}
i\,G_M^{(1)}(p)^{-1}= (\omega^2 - \vec k^2) \left[1+  \frac{|h|^2}{16}\frac{1}{\sqrt{ \vec k^2-\omega^2}}\right]\,.
\ee

\begin{figure}[h!]
\begin{center}
\includegraphics[width=0.5\textwidth]{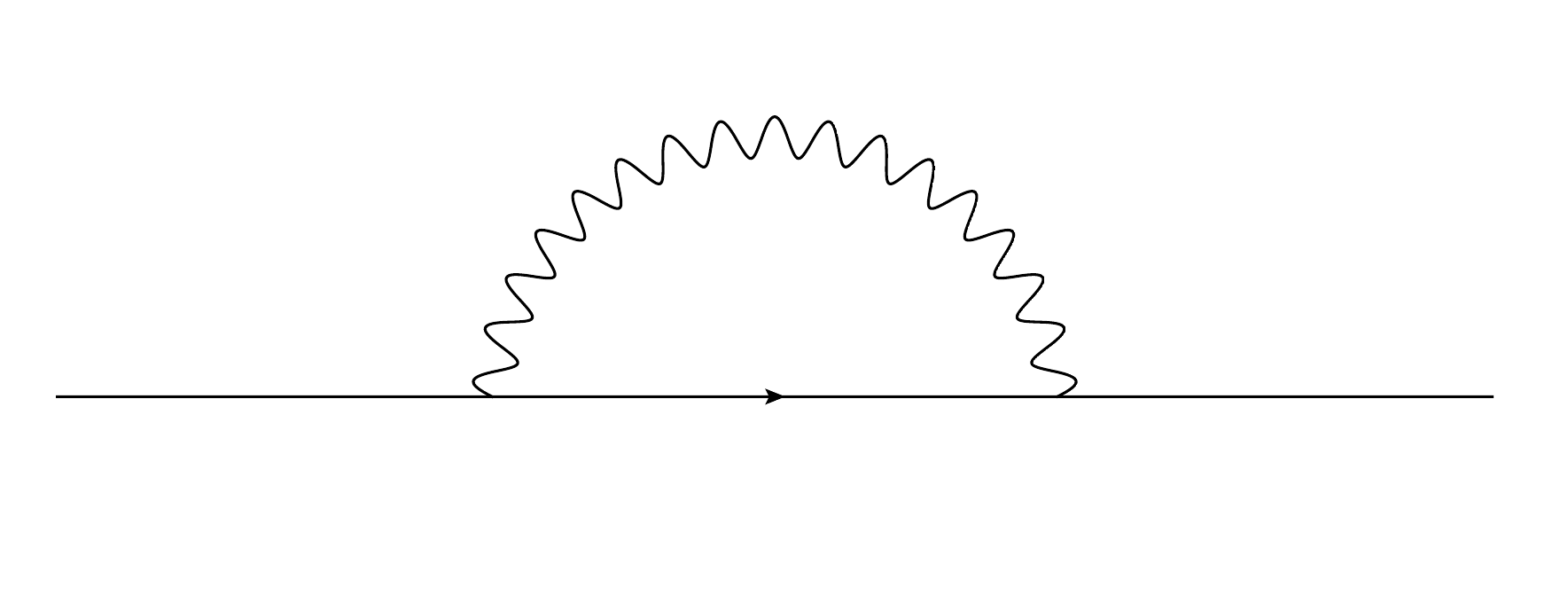}
\end{center}
\caption{\small{One-loop self-energy diagram for the charged fermion fields.}}\label{fig:fermionself}
\end{figure}

Similarly, for the self-energies of the charged fermions we obtain from the diagrams in Figure \ref{fig:fermionself}
\be
\Sigma_{\psi_\pm}(p)= |h|^2 \int \frac{d^3q}{(2\pi)^3}\, G_{\psi_\mp}^{(0)}(q) \,G_M^{(0)}(p+q)= |h|^2 \int \frac{d^3q}{(2\pi)^3}\,\frac{i}{\slashed q \mp \hat{\slashed A}}\,\frac{i}{(p+q)^2}\,.
\ee
After a redefinition of the loop momentum, this one-loop diagram coincides with the relativistic one, but with a shifted external momentum $p \pm \hat{\slashed A}$.
The inverse one loop propagator then reads
\be\label{eq:Gpsi}
iG^{(1)}_{\psi_\pm}(p)^{-1}= (\slashed p \pm \hat{\slashed A})\left[1+  \frac{|h|^2}{32} \frac{1}{\sqrt{\vec p^2-(\omega\pm \mu)^2 }} \right]\,.
\ee

In both cases we find nontrivial contributions to the tree level kinetic term. These modify the IR scaling dimensions and lead to non Fermi liquid behavior. For the boson, the wavefunction renormalization $Z_M$ becomes singular on mass-shell; this comes from the cancellation of the chemical potential contributions of $\psi_+$ and $\psi_-$ inside the loop. For the charged fermions, the wavefunction renormalization $Z_{\psi_\pm}$ becomes singular on the Fermi surface, signaling the fact that we have integrated out light degrees of freedom.

An important simplification in this theory is that there is no one loop renormalization of the coupling $h$, a consequence of having an interaction of the form $V \supset h M \psi_+ \psi_-$. Therefore, the running of the physical coupling at one loop is determined solely in terms of wavefunction renormalization. This property is inherited from the supersymmetric theory, where holomorphic superpotential interactions are not renormalized in perturbation theory.

\subsection{Comments on infrared dynamics}

In the previous section we computed the one loop corrections to the magnetic theory, where a gapless boson interacts with Fermi surfaces for the vortices $\psi_\pm$. At energy scales $E \sim |h|^2$ the theory becomes strongly coupled and perturbation theory cannot be trusted. This is also the regime where the duality between the electric and magnetic descriptions holds. We will now discuss some aspects of the long distance dynamics and ways to control it, leaving a more detailed analysis to the future.  Very similar systems have been analyzed in detail, with differing assumptions, in the 
recent works \cite{lee1,MS,MIT,Fitzpatrick:2013mja,lee2}.

The first possibility is to generalize the model to include $N$ flavors $\psi_\pm$. Taking $N \gg 1$, it may be possible to have a controlled $1/N$ expansion that leads to a fixed point,
or at least an approximate fixed point which governs the theory over a certain energy range. To see this, it is useful to introduce the dimensionless physical coupling
\be
\bar h^2 = Z_M Z_{\psi_+} Z_{\psi_-}\,\frac{|h|^2}{E}
\ee
where $E$ is an energy scale and $Z_i$ are the wavefunction renormalization factors computed in \S \ref{subsec:oneloop}.
At large $N$ the one loop self-energy corrections to the fermions are subleading, so if we are far enough from the Fermi surface, so at a momentum-scale $k$ we have
\be
\bar h^2 \approx \, \left(1+ N \frac{|h|^2}{16}\frac{1}{|k|}\right)^{-1} \,\frac{|h|^2}{|k|}\,.
\ee
At low energies, this gives a fixed point $\bar h^2 \approx 16/N$. This fixed point is similar to those discussed in ~\cite{Appelquist:1981sf}. One important difference in our setup is the presence of a Fermi surface; in particular, close enough to the Fermi surface the $1/N$ expansion breaks down due to the singularity in (\ref{eq:Gpsi}). The infrared singularities from the Fermi surface excitations may also affect the theory through higher loops, as in~\cite{lee1}.

It would be interesting to study the large $N$ dynamics of this theory in more detail. Our previous arguments suggest the possibility of a fixed point for the gapless boson interacting with the Fermi surfaces, or at least an approximate fixed point which governs the theory over some range of
momenta. Such a phase of matter is believed by some to be relevant for high $T_c$ superconductors, where the gapless boson can appear as a gauge redundancy from the spinon-holon separation.  For this reason, such theories have been intensely studied in the past, using large $N$ to extract results at long distance~\cite{Polchinski:1993ii, wen, millis}. More recently~\cite{lee1} argued that IR divergences from the Fermi surface invalidate the large $N$ expansion. It would be interesting to understand whether the arguments of that work apply also to our model, whose matter content and interactions are different from~\cite{Polchinski:1993ii}.

Another option in order to have a perturbative expansion is to consider the theory near its upper critical dimension, $d=4 - \epsilon$, with small $\epsilon$. In this limit the boson and fermion self-energy have logarithmically divergent contributions and hence nonvanishing anomalous dimensions. Balancing the one loop contributions to the beta function $\beta_h$ against the classical running from $[h]=\epsilon/2$ yields a fixed point with $|h|^2 \sim \epsilon^2$, and anomalous dimensions $\gamma \sim \epsilon$ --a generalization of the Wilson-Fisher fixed point. This leads to a non Fermi liquid under perturbative control, which we hope to analyze in detail in a future work.  A similar theory was recently discussed in detail in \cite{Fitzpatrick:2013mja}.

\section{Conclusions}\label{sec:concl}

It has been clear for many years that supersymmetric field theories provide a valuable playground
for understanding new phenomena in strongly coupled quantum field theory.  To date, this has
largely been exploited to study the dynamics of Lorentz-invariant vacua of such theories.  In this
paper and in \cite{Hook:2013yda}, we have started to explore the lessons that supersymmetric
dualities might hold for the physics of strongly coupled systems at finite charge density.

By examining the simplest mirror pairs of 3d supersymmetric abelian gauge theories, doped with a
chemical potential for the $U(1)_J$ global charge characteristic of such theories, we were able
to exhibit examples of several interesting phenomena.  These include the emergence of 
itinerant fermions at a Fermi surface from a UV theory with fermions localized at defects; 
emergent gauge fields and spin-charge separation; and dual descriptions of a non-Fermi liquid arising from fermions at a Fermi surface coupled
to a critical boson.  

It is clear that the phase diagram of abelian gauge theories doped by impurities is very rich, and here we have explored just the simplest possibilities. In particular, it will be of interest to add a finite density of fermions to the electric description that contains the lattice of magnetic impurities. Furthermore, it would be useful to extend these results to models with Chern-Simons interactions (see e.g.~\cite{Tong:2000ky,Jafferis:2008em}), with less SUSY as in~\cite{Gukov:2002es} or in other dimensions. See for instance~\cite{Harnik:2003ke,Cherman:2013rla} for an analysis of 4d supersymmetric QCD and QED with nonzero chemical potential.

The results found here suggest that the framework of supersymmetric dualities might be
useful in understanding several problems of condensed matter physics.  For instance,
it seems likely that one can find phase transitions with Fermi surface reconstruction, with 
the Luttinger count at the Fermi surface increasing due to contributions from magnetic defects
-- a problem whose physical interest is well described in \cite{Senthil}.  This is also a natural
setting to search for soluble examples of Kondo lattice models, or more general non-Fermi liquids.
It is also quite plausible that in addition to obtaining lessons for field theory at finite density
by perturbing known mirror dualities, one may be able to find intrinsically new dualities governing
the non-relativistic fixed points of doped supersymmetric field theories.   These constitute
promising problems for future research.

\section*{Acknowledgments}

We would like to thank A.L. Fitzpatrick, S. Hartnoll, J. Kaplan, S. Kivelson, M. Mulligan, S. Raghu and 
E. Silverstein for many discussions about related issues in field theory and condensed matter physics. We also thank A.L. Fitzpatrick, S.S. Lee, J. McGreevy, and M. Mulligan  for comments on the draft.
 A.H.~is supported by the Department of Energy under contract DE-FG02-90ER40542.
The research of S.K. and G.T.~is supported in part by the National Science Foundation under grant no.~PHY-0756174.  S.K. is also supported by the Department of Energy under contract
DE-AC02-76SF00515, and the John Templeton Foundation. H.W.~ is supported by a Stanford
Graduate Fellowship.

\bibliographystyle{JHEP}
\renewcommand{\refname}{Bibliography}
\addcontentsline{toc}{section}{Bibliography}
\providecommand{\href}[2]{#2}\begingroup\raggedright

\end{document}